\documentclass[aps,twocolumn,secnumarabic,amsmath,amssymb,
nofootinbib]{revtex4-2}

\usepackage{graphics}      
\usepackage{graphicx}      
\usepackage{longtable}     
\usepackage{bm}            
\usepackage{url}           
\usepackage{hyperref}
\hypersetup{
	breaklinks=true,
	colorlinks=true,
	linkcolor=blue,
	filecolor=magenta,      
	urlcolor=cyan,
	pdfpagemode=FullScreen,
}

\begin{document}
\title{Particle-in-cell simulations of laser crossbeam energy transfer via magnetized ion-acoustic wave}
\author         {Yuan Shi}
\email          {Yuan.Shi@colorado.edu}
\affiliation    {Department of Physics, Center for Integrated Plasma Studies, University of Colorado Boulder, Boulder, 	CO 80309, USA}

\author         {John D. Moody}
\affiliation    {Lawrence Livermore National Laboratory, Livermore, California 94550, USA}

\date{\today}

\begin{abstract}
	Large magnetic fields, either imposed externally or produced spontaneously, are often present in laser-driven high-energy-density systems. In addition to changing plasma conditions, magnetic fields also directly modify laser-plasma interactions (LPI) by changing participating waves and their nonlonear interactions. In this paper, we use two-dimensional particle-in-cell (PIC) simulations to investigate how magnetic fields directly affect crossbeam energy transfer (CBET) from a pump to a seed laser beam, when the transfer is mediated by the ion-acoustic wave (IAW) quasimode. Our simulations are performed in the parameter space where CBET is the dominant process, and in a linear regime where pump depletion, distribution function evolution, and secondary instabilities are insignificant. We use a Fourier filter to separate out the seed signal, and project the seed fields to two electromagnetic eigenmodes, which become nondegenerate in magnetized plasmas. By comparing the seed energy before CBET occurs and after CBET reaches quasi-steady state, we extract CBET energy gains of both eigenmodes for lasers that are initially linearly polarized. 
	Our simulations reveal that starting from a few MG fields, the two eigenmodes have different gains, and magnetization alters how the gains depend on laser detuning. The overall gain decreases with magnetization when the laser polarizations are initially parallel, while a nonzero gain becomes allowed when the laser polarizations are initially orthogonal. These findings qualitatively agree with theoretical expectations. 
\end{abstract}

\maketitle

\section{Introduction}
When laser-driven inertial confinement fusion (ICF) experiments are pre-magnetized using external coils, the additional magnetic confinement increases ion temperature and fusion yield \cite{Chang11,Hohenberger12,Hansen20,Moody22,Peebles23}. Beyond existing experiments, simulations of magnetized ICF implosions \cite{Perkins17,Walsh22,Strozzi24,Walsh25} suggest that magnetic fields may also stabilize Rayleigh–Taylor instabilities and reduce mix, with plausible inhibition of fusion burn propagation \cite{Appelbe21, Neill25}. When combining magnetization effects optimally, achieving more robust fusion ignition at reduced driver laser energy and relaxed target smoothness requirements may be possible.  

Even when magnetic fields are not imposed externally, ICF experiments, and more broadly speaking, high-energy density (HED) experiments, tend to be spontaneously magnetized. A well-studied source of spontaneous magnetization is the Biermann-battery effect \cite{Stamper91, Sherlock20, Walsh21, Campbell22}, where non-parallel density and temperature gradients drive the growth of magnetic fields. Additionally, the Poynting-Robertson effect \cite{Cattani66,Munirov17} and kinetic effects \cite{Liu24} can also drive seed magnetic fields, which may be subsequently amplified by flux compression \cite{Gotchev09, Knauer10}. Near laser entrance holes of indirect-drive implosions, where fields are compressed by plasma inflow, $\sim 10^2$ T spontaneous fields are common \cite{Farmer17}. Moreover, near the center of a pre-magnetized capsule, $\sim 10^3$ T fields have been inferred in experiments using nuclear diagnostics \cite{Sio21}.

In magnetized HED experiments, the field modifications to the laser-plasma interactions (LPI) is not well understood \cite{Shi18a}. To estimate when magnetization directly affects LPI, as opposed to indirectly through changing plasma conditions, we can compare characteristic frequency scales. In \mbox{$\sim 1$ MG} fields, the electron gyro frequency $\omega_{ce}$ becomes comparable to typical frequencies of ion-acoustic waves (IAW) that mediate Brillouin scattering and crossbeam energy transfer (CBET). In \mbox{$\sim 10$ MG} fields, $\omega_{ce}$ becomes comparable to the frequency of electron plasma waves (EPW) that mediate Raman scattering and two-plasmon decay. In even stronger fields, $\omega_{ce}$ becomes comparable to the laser frequency, where drastic changes to LPI occur \cite{Edwards19, Manzo22}. Moreover, when the plasma is magnetized, a zoo of additional plasma waves, which have no analogue in unmagnetized plasmas, can mediate additional resonant interactions \cite{shi2018plasma, Shi21, Shi23}. 
Existing studies of magnetized LPI (MagLPI) mostly focus on how magnetic fields modify Raman- and Brillouin-type interactions \cite{Boyd_Rankin_1985}, through modifying plasma conditions and hot electrons \cite{Yao23}, changing collisionless wave damping \cite{Winjum18}, and splitting polarizations of electromagnetic eigenmodes \cite{Nabil98}.

Although the study of MagLPI is scarce, magnetized three- and four-wave interactions \cite{Galloway_Kim_1971, Boyd78} and parametric instabilities \cite{Liu1986143, Cohen87} , which are the more general overarching phenomena, have been investigated in other contexts. 
For example, in magnetic confinement fusion, antennas mounted on vacuum chamber walls launch large amplitude radio-frequency waves and excite parametric instabilities \cite{Porkolab_1978}. In this near perpendicular geometry, the pump wave can excite parametric instabilities and decay to upper- and lower-hybrid waves \cite{Grebogi-Liu-80, Hansen_2017}, as well as Bernstein waves \cite{Platzman68, Stenflo81}, which can subsequently cascade to other waves \cite{kasymov1985upper}.
As another example, in astrophysical plasmas, waves often propagate nearly parallel to the magnetic field. In this context, the pump wave is usually an Alfvén wave whose frequency is below ion cyclotron frequencies. The pump Alfvén wave can couple with sound waves and other Alfvén type waves \cite{Hasegawa-Liu_76, Wong-Goldstein86, Viñas_Goldstein_1991}.
In addition to these examples, a wealth of nonlinear processes occur in magnetized plasmas. 
This paper is dedicated to the 85th birthday of Prof. Lennart Stenflo, who pioneered the study of magnetized wave-wave interactions starting from the late 60s and early 70s \cite{Sjolund67,sjolund1967parametric, stenflo1970solution,stenflo1971non}. 
Together with his collaborators, Prof. Stenflo has developed cold-fluid \cite{stenflo1973three}, warm-fluid \cite{Larsson1973three}, kinetic \cite{Stenflo70}, and collisional \cite{stenflo1970effect,shukla2002modulational} theories of magnetized three-wave interactions. 
They derived fluctuation spectrum enhanced by pump waves \cite{Larsson76, Stenflo_2004}, investigated effects of nonuniformity \cite{Kaufman_1979}, plasma boundary \cite{linden1982three, Lindgren_Stenflo_Kostov_Zhelyazkov_1985}, gravity potential \cite{axelsson1996nonlinear,STENFLO_SHUKLA_2009}, and nonstationary turbulence \cite{vladimirov1997three}. 
They studied many special cases \cite{brodin1989three,brodin2007nonlinear, Brodin2016new}, as well as made numerous attempts to simplify general expression of the coupling coefficients \cite{Larsson1976parametric, Stenflo94, Stenflo2006three, Brodin12}.
Their theories have been applied to ionospheric experiments \cite{Stenflo_1990}, free-electron lasers \cite{Stenflo1981self}, THz radiation generation \cite{brodin2014wave}, and quantum plasmas \cite{Stenflo2009wave}.

In our companion paper \cite{Moody25}, we develop a Vlasov theory of magnetized CBET (MagCBET) using a low-frequency approximation of magnetized ponderomotive force. 
In laser driven implosions, CBET is an important LPI process where energy is transferred from one laser beam to another \cite{Randall81}. In unmagnetized plasmas \cite{Michel23}, CBET is mediated by IAW quasimodes, and the resonant conditions are satisfied either due to a plasma flow, in which laser beams are Doppler shifted by different amounts, or due to an intentional laser wavelength detuning $\Delta\lambda$ at the lens. In direct-drive implosions, CBET often leads to loss of coupling and symmetry, when incoming lasers are consumed to amplify backscattered beams \cite{Seka08,Froula12,Marozas18}. In indirect-drive implosions, CBET is actively used to control the implosion symmetry, whereby energy is redistributed between polar and equatorial beams to achieve spherical implosions \cite{Michel11,Moody12,Pickworth20}. 
Our MagCBET theory suggest three new physical effects at fixed plasma conditions. 
First, the coupling between initially parallel polarized lasers is reduced by magnetization. 
Second, orthognally polarized lasers, which do not couple in unmagnetized plasmas, become coupled due to magnetization.
Third, the gain curve, namely, the CBET gain as a function of $\Delta\lambda$, is altered by magnetization, with additional resonances away from the IAW resonance. 
When plasma is magnetized, the two electromagnetic eigenmodes are nondegenerate and have different gains. The total gain of a linearly polrized seed laser is distributed between the eigenmode gains.

In this paper, we verify these theoretical findings on MagCBET using particle-in-cell (PIC) simulations in the plasma rest frame and we focuse on the IAW resonance. The simulations are performed using the EPOCH code \cite{Arber:2015hc}, which allows for an arbitrary background magnetic field and is fully relativistic and electromagnetic. The PIC simulations are two dimensional in space with all three field components and velocity directions (2D-3V). To compare with our Vlasov theory, we turn off collisions, which may nevertheless have effects, especially in lower temperature experiments \cite{Turnbull20,Hansen22}. To avoid complications due to additional nonlinear effects \cite{Seaton22,Yin23}, we choose simulation parameters to stay within the linear regime of CBET, namely, when pump depletion and distribution function evolution are insignificant and when secondary instabilities are weak. In this linear regime, our simulations qualitatively confirm our theoretical findings. However, quantitative differences are observed, suggesting additional physics beyond our Vlasov theory. 

The paper is organized as follows. In Sec.~\ref{sec:setup}, we describe our simulation setup and our considerations for choosing simulation parameters. In Sec.~\ref{sec:data}, we describe how we extract CBET gain from raw PIC data. In Sec.~\ref{sec:results}, we present representative results of our simulations, and additional results can be found in \cite{Shi25data}. A summary and discussion is given in Sec.~\ref{sec:discussion}.

\section{Simulation setup\label{sec:setup}}
\begin{figure}[t]
	\centering
	\includegraphics[width=0.75\linewidth]{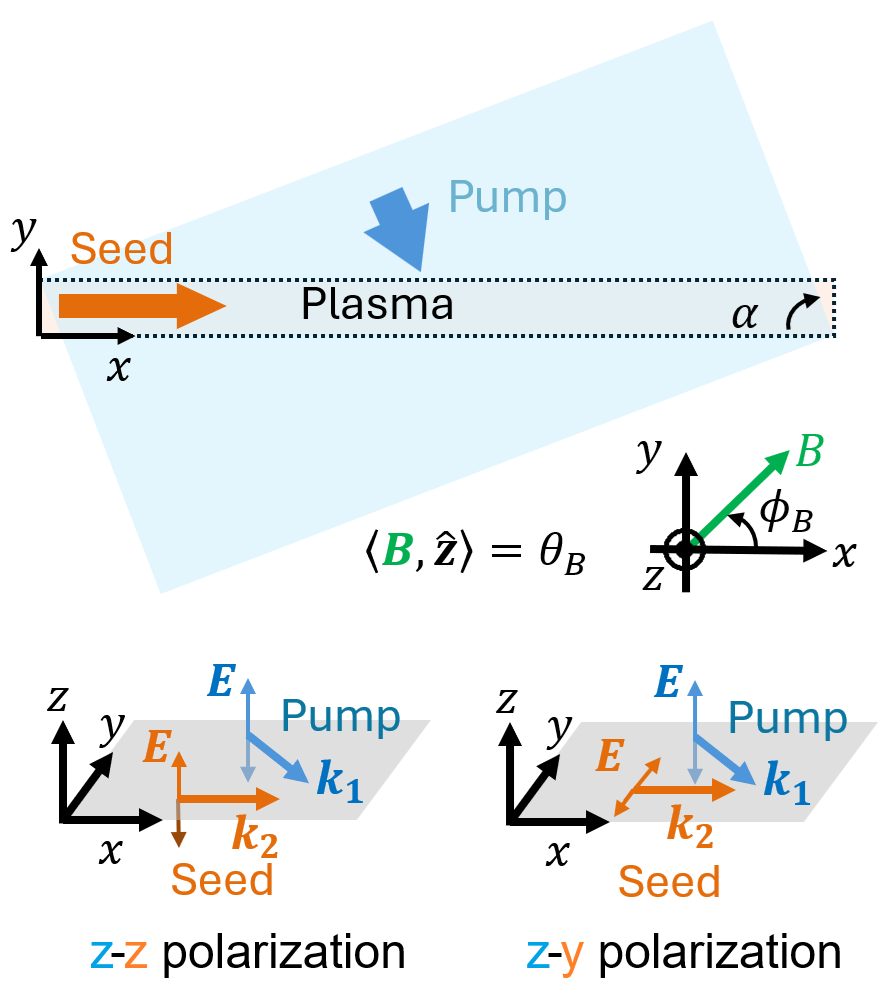} 
	\caption{Setup of 2D-3V PIC simulations, where the simulation domain (dashed rectangle) is centered around the seed laser and magnetic field is at oblique angles. For $z$-$z$ polarization, the pump and seed lasers have initially parallel polarizations, while for $z$-$y$, their polarizations are initially orthogonal.
	}
	\label{fig:Setup}
\end{figure}

To reduce computational cost, we use an elongated rectangular simulation domain centered around the seed laser beam, as shown in Fig.~\ref{fig:Setup}. The domain is within $0<x<L_x$ and $-L_y/2<y<L_y/2$, where $L_x=500\lambda_0$, $L_y=10\lambda_0$, and $\lambda_0=0.351\,\mu$m is the pump laser wavelength in vacuum. 
A domain of length $L$ gives a finite wavevector resolution $\Delta k=2\pi/L$.  
The seed laser enters from the $-x$ domain boundary, propagates with unit wavevector $\hat{\mathbf{k}}=\hat{\mathbf{x}}$, and then exits the $+x$ domain boundary. All boundaries use Convolutional Perfectly Matched Layer (CPML) to realize absorbing boundary conditions for electromagnetic waves \cite{Taflove05}. 
In free space, the seed laser is a Gaussian beam, focused at the center of the domain with a half divergence angle of $4^\circ$. 
However, due to a finite apeture, the seed laser is diffracted, and the Rayleigh angle is $1.22\lambda/L_y\approx7^\circ$, where $\lambda\approx\lambda_0$ is the seed laser wavelength in vacuum. Since the Rayleigh angle is larger than the Gaussian focusing angle, the seed beam is slightly divergent, leading to a small loss from $\pm y$ domain boundaries. A larger $L_y$ can reduce the loss but will increase the computational cost, so is not used here. 
Additionally, because the domain is filled with a plasma, the seed laser undergoes refraction when entering and exiting the domain. Nevertheless, because the plasma density we use is much lower than the critical density of the laser, the refraction angle is miniscule. 
The overall effect, with diffraction overcoming focusing, is that the seed laser is approximately a plane wave, with a Gaussian intensity envelope along the $y$ direction and an approximately $y$-independent normalized envelope along $x$.

To observe CBET, which changes the seed envelope along the $x$ direction, we turn on a pump laser and fill the domain with a magnetized plasma. The pump laser is a plane wave with unit wavevector $\hat{\mathbf{k}}_0=\hat{\mathbf{x}}\cos\alpha - \hat{\mathbf{y}} \sin\alpha$, where $\alpha=30^\circ$ is the crossing angle between the pump and seed lasers. The pump laser enters from both the $-x$ and $+y$ boundaries, and exits from the $+x$ and $-y$ boundaries. Because of a large incidence angle to the CPML, which is designed to absorb waves with near-normal incidence, about 5\% pump energy is reflected from the exit boundaries when the pump laser is linearly polarized along the $\hat{\mathbf{z}}$ direction at the entrance boundaries. 
At the entrance boundaries, we fix the pump laser to $\hat{\mathbf{z}}$ polarization, and simulate two seed laser polarizations: In the $z$-$z$ polarization setup, the seed laser is also linearly polarized along the $\hat{\mathbf{z}}$ direction at the entrance boundary, so the pump and seed lasers have initially parallel polarizations. On the other hand, in the $z$-$y$ polarization setup, the seed laser is linearly polarized along the $\hat{\mathbf{y}}$ direction at the entrance boundary, so the pump and seed lasers have initially orthogonal polarizations. 
As the lasers propagate through the magnetized plasma, their polarization vectors rotate and gain ellipticity. 
The entire domain is filled with a plasma of density $n_e=3.5\times 10^{20}\;\text{cm}^{-3}$. We use a fully ionized carbon plasma with initially Maxwellian distribution functions at temperatures $T_e=4T_i=$ \mbox{0.8 keV}. The plasma is embedded in a uniform magnetic field. We specify the direction of $\mathbf{B}$ in standard spherical coordinates, with polar angle $\theta_B$ and azimuthal angle $\phi_B$.

We choose simulation parameters such that linear CBET is the dominant process in our PIC simulations, which also capture other competing processes. 
Notice that the plasma has a sheath field $E_S\sim\phi_S/\lambda_D\sim $\mbox{$10^{11}$ V/m}, where $\phi_S\sim k_B T/e$ is the sheath potential, and $\lambda_D$ is the Debye length. The sheath field is comparable to the laser field $E_L=\sqrt{2I/\epsilon_0 c}\sim 10^{11}\sqrt{I_{15}}$ V/m, where $I_{15}$ is the laser intensity in units of $10^{15}\,\text{W/cm}^2$. Because of the small $L_y$, plasma expansion due to the sheath field has an appreciable effect unless $I_{15}=O(1)$. We choose pump and seed laser intensities to be $I_p=I_s=0.5\times 10^{15}\,\text{W/cm}^2$ at the entering boundaries.  
When we double $I_s$, noticeable spontaneous backscattering is observed for the seed laser. When we halve $I_s$ or double $I_p$, instabilities seeded by PIC noise starts to compete. When we halve $I_p$, a larger $L_x$ is needed to achieve $O(1)$ gain for CBET. However, a larger $L_x$ also allows more growth of PIC-noise seeded instabilities, and is thus unfavorable. 
We scan numerical resolutions and find that using 20 particles per cell and 20 cells per pump wavelength is usually sufficient for achieving convergence. 

\section{Data analysis \label{sec:data}}
From raw PIC data, we extract the seed laser fields using a Fourier filter. 
Suppose $E(x,y)$ is the data of a field component, we take 2D discrete Fourier transform to find $\hat{E}(k_x, k_y)$. Apart from peaks corresponding to the seed laser, the complex-valued $\hat{E}$ includes peaks due to the incident pump laser, the reflected pump, and noise-seeded instabilities. 
We pick out the seed laser peaks using a bandpass filter, which applies a weight $w(k_x, k_y)$ to $\hat{E}$. 
The weight is $w=1$ within two square windows centered at $k_x=\pm k_0\equiv\pm 2\pi/\lambda_0$ and $k_y=0$. The windows have widths $\Delta k_x=5\% k_0$ and $\Delta k_y=2\pi/L_y$. Outside the windows, the weight has a Gaussian profile, which falls off from 1 to 0 with standard deviations $\sigma_{k_x} = \Delta k_x$ and $\sigma_{k_y} = \Delta k_y$.
This Gaussian bandpass filter removes most waves related to the pump laser, as well as the sheath electric field and background magnetic field, while retaining waves related to the seed laser. 
After filtering, we apply inverse discrete 2D Fourier transform on $w\hat{E}$ to reconstruct $E_s(x,y)$, the real-valued seed field in the configuration space. 
Because the seed field is approximately a plane wave, we reduce the 2D data to a 1D data by integrating along the $y$ direction. 
The 1D seed field is $\tilde{E}(x) = \int_{-h}^{h} dy\; E_s(x,y)$, where $h=f_yL_y/2$ and we vary the fraction $0<f_y<1$ to estimate uncertainties. 

Using 1D seed fields $\tilde{E}_y$ and $\tilde{E}_z$, we synthesize fields of two underlying eigenmodes, which are non-degenerate in magnetized plasma.
When the laser frequency is much larger than the electron cyclotron frequency, the two eigenmodes are approximately transverse. 
Because the seed propagates along the $x$ direction, 
complex unit polarization vectors of the two orthogonal eigenmodes can be parameterized as 
\begin{eqnarray}
	\hat{\mathbf{e}}_1 &=& (0, \cos\phi, e^{i\psi}\sin\phi),\\
	\hat{\mathbf{e}}_2 &=& (0, \sin\phi, -e^{i\psi}\cos\phi),
\end{eqnarray}
where the polarization angles $\phi$ and $\psi$ are computed using a warm-fluid theory \cite{PhysRevE.99.063212, Shi22}. At $B=0$, the eigenmodes $\hat{\mathbf{e}}_1$ and $\hat{\mathbf{e}}_2$ are right- and left-handed circularly polarized with respect to $\mathbf{k}$. 
At $B\ne 0$, we identify $\hat{\mathbf{e}}_1$ with the X mode, and $\hat{\mathbf{e}}_2$ with the O mode. These two mode branches reduce to the usual X and O modes when the wave propagates perpendicular to the magnetic field, where the X-mode electric field is perpendicular to the background magnetic field $\mathbf{B}$ while the O-mode electric field is parallel to $\mathbf{B}$. 
On the other hand, when the wave propates parallel to $\mathbf{B}$, the X-mode branch becomes the R wave, which is right-handed circularly polarized with respect to $\mathbf{B}$, while the O-mode branch becomes the L wave, which is left-handed circularly polarized with respect to $\mathbf{B}$.
From the unit polarization vectors, the complex electric field at a snapshot in time is $\mathbf{E}_C = \mathcal{E}_1\hat{\mathbf{e}}_1 e^{ik_1x} + \mathcal{E}_2\hat{\mathbf{e}}_2 e^{ik_2x}$. Denoting the complex envelope as $\mathcal{E}=A e^{i\alpha}$, where $A$ and $\alpha$ are slowly varying amplitude and phase of the envelope, the real field $\mathbf{E} = \text{Re}\, \mathbf{E}_C$ has components
\begin{eqnarray}
	\tilde{E}_y &=& A_1\cos\phi\cos\theta_1 + A_2\sin\phi\cos\theta_2,\\
	\tilde{E}_z &=& A_1\sin\phi\cos(\theta_1+\psi) - A_2\cos\phi\cos(\theta_2+\psi),
\end{eqnarray}
where $\theta_j\equiv k_j x+\alpha_j$ and $j=1,2$.
If the simulation domain is long enough to resolve the difference of the two eigenmode wavevectors, namely, when $\Delta k = (k_1-k_2)/2\gg 1/L_x$, Fourier transforms of simulation data will show two distinct peaks, and the two eigenmodes are separable using a Fourier filter. 
However, our simulations are in the regime $\Delta k \sim 1/L_x$, where the two Fourier peaks merge. In this regime, a different scheme is needed to separate the two eigenmodes: 
We approximate the two modes as having a common $\bar{k}=(k_1+k_2)/2$. 
We then reconstruct the eigenmodes from linear combinations of $\tilde{E}_y$ and $\tilde{E}_z$. Because $\tilde{E}_y$ and $\tilde{E}_z$ are not in phase, to linearly combine them, we shift $\tilde{E}_z\rightarrow \tilde{E}'_z(x)=\tilde{E}_z(x-\Delta x)$, where $\bar{k}\Delta x = \psi$. After the shift, the field compomnents become in-phase, so the eigenmode fields can be reconstructed by
\begin{eqnarray}
	\tilde{E}_1 &\equiv& A_1\cos\theta_1 \approx \cos\phi\, \tilde{E}_y + \sin\phi\, \tilde{E}'_z,\\
	\tilde{E}_2 &\equiv& A_2\cos\theta_2 \approx \sin\phi\, \tilde{E}_y - \cos\phi\, \tilde{E}'_z.
\end{eqnarray}
While $\tilde{E}_y$ and $\tilde{E}_z$ subject to effects like Faraday rotation, envelopes of the eigenmode fields $\tilde{E}_1$ and $\tilde{E}_2$ change along $x$ primarily due to CBET.

Using eigenmode fields $\tilde{E}_1$ and $\tilde{E}_2$, we estimate CBET energy gain when simulations have reached quasi steady state. For eigenmode $j$, the energy gain at time $t$ is defined as 
\begin{equation}
	g_j(t)=\ln\frac{U_j(t)}{U_j(t_0)},
\end{equation}
where the electric-field areal energy density is $U_j = \frac{\epsilon_0}{2}\int_0^{L_x} dx\; \tilde{E}_j^2$, and $t_0$ is the time when the seed laser has completely filled the simulation domain but the pump laser has just begun to enter. 
In our setup, the seed laser begins to enter at $t=0$, so $t_0\approx $ \mbox{0.6 ps}, which is the time for the front the seed laser to reach the other end of the simulation box. A quasi steady state is reached around $t_1=$ \mbox{3 ps}, and $U_j$ remains approximately the same around $t_2=$ \mbox{3.6 ps}. In PIC simulations, because the distribution functions evolve on a slower time scale, there is no true steady state, and we use both $g_j(t_1)$ and $g_j(t_2)$ to estimate uncertainties. 
Similar to eigenmode energy gains, we also compute the total gain, which is defined using the total areal energy density $U=U_1+U_2 = U_y+U_z$. We verify that these two ways of computing $U$ give consistent results. 
The energy gain $g$ is closely related to the amplitude gain $G$ in CBET theory, in which one finds the imaginary part of the wavevector at a given real frequency. Suppose the complex wavevector is $k=k_R+ik_I$, then the complex field is $E_c\propto e^{ikx}=e^{ik_Rx} e^{-k_Ix}$. When $k_I<0$, the envelope grows exponentially in $+x$. The envelope gain is defined as $G=-2k_IL_x$. Integrating exponential envelope in space, the energy gain is $g=\ln F(G)$, which is related to the amplitude gain $G$ by 
\begin{equation}
	\label{eq:gain_energy_amplitude}
	F(G)=(e^G-1)/G.
\end{equation} 
In the limit $G\rightarrow 0$, the energy gain is $g\simeq G/2$. In our PIC simulations, $g=O(1)$ is in the small gain regime. In this regime, pump depletion is insignificant, partly also because the pump freshly enters and exits from the transverse domain boundaries with only a small transit length. However, we notice that the seed eigenmode envelopes have noticeable differences from exponential profiles, possibly due to competing instabilities in PIC simulations, as well as artifacts from our data analysis method, which introduces distortions especially near boundary points due to Fourier filtering. We report energy gain, which is well defined regardless of the profiles. The energy gain is also what is typically measured in experiments.

To compare theory with PIC simulations and experiments, we also take into account finite plasma size effects that causes blurring of the gain curve, namely, $g$ as a function of the detuning $\Delta\lambda = \lambda-\lambda_0$.
To understand the blurring, suppose a laser enters from vacuum with frequency $\omega$. In an infinite plasma, the laser wavevector would be $k$ from the linear dispersion relation. However, in a plasma of a finite size $L_x$, the modes are discrete with $\Delta k = 2\pi/L_x$. When $k$ does not match the wavevector of one discrete mode, the laser excites multiple modes. When $\Delta k\ll k$, the laser approximately excites two adjacent modes with $k_-<k<k_+$. The complex laser field is a linear superposition $E_C\approx \mathcal{E}(f_+ e^{ik_+ x} + f_-e^{ik_- x})$, where the amplitudes are partitioned as
\begin{eqnarray}
	f_+ =  \frac{k-k_-}{\Delta k}, \quad
	f_- =  \frac{k_+-k}{\Delta k},
\end{eqnarray}
with $f_+ + f_- = 1$. 
In particular, when $k=k_-$, the partition is $f_+=0$ and $f_-=1$, so only the $k_-$ mode is excited, and similarly for $k=k_+$. 
Using Eq.~(\ref{eq:gain_energy_amplitude}), the areal energy density of the superimposed excitation is 
\begin{eqnarray}
	\label{eq:energy_two_modes}
	\nonumber
	U&=&\frac{\epsilon_0}{4}|\mathcal{E}|^2 L_x \Big[ f_+^2 F(G_+) + f_-^2F(G_-) \\
	&+& 2f_-f_+ F\Big(\frac{G_+ + G_-}{2}\Big)\langle\cos(\delta k \, x)\rangle \Big],
\end{eqnarray}
where $\delta k = k_R^+ - k_R^-$ is on the order of $\Delta k$, but is modified by nonlinear dispersion relation due to CBET.
The average is defined as $\langle\dots\rangle\equiv\frac{1}{L_x}\int_0^{L_x} dx\dots$. Because $\delta k=O(1/L_x)$, the average $\langle\dots\rangle$ is not zero, but is rather sensitive to values at the boundary points. Nevertheless, the average is bounded $|\langle\dots\rangle|\le \frac{2}{L_x \delta k}$. 
Because two discrete modes are excited, the total gain is different from their respective single-mode gains due to an averaging effect. Therefore, compared to the gain curve $g(\Delta\lambda)$ in the continuum, $g(\Delta\lambda)$ in a finite-size plasma is broadened. 
This finite-size broadening effect can be estimated using Eq.~(\ref{eq:energy_two_modes}) when comparing numeric $g$ with analytic $G_\pm$.

To map out CBET gain curves, we fix the pump vacuum wavelength $\lambda_0$, and scan the seed vacuum wavelength $\lambda$.  
In our simulations, the seed is launched from the boundary using an antenna driven at frequency $\omega = 2\pi c/\lambda$. Because of a fine time resolution, small $\Delta\omega=\omega_0-\omega$ is resolved, even though the spatial domian size can only resolve $\Delta k = 2\pi/L_x$, which limits $\Delta \lambda = \lambda\lambda_0/L_x=\lambda/500\approx7 \mathring{A}$.
As mentioned earlier, increasing $L_x$, which is already on par with interaction lengths typically used in experiments, is unfavorable due to competing PIC processes. 
In the following section, we report error bars for $g$ including two effects. First, due to the competition of CBET with other PIC processes, the gain is not in a true stready state. To estimate the uncertainty due to slow deviations away from the steady state, we measure the gains at two simulation time $t_1$ and $t_2$. 
Second, due to the competition between Gaussian focusing and finite-apeture diffraction, the seed laser is not a plane wave. To estimate the uncertainty due to the fact that the wave field $\tilde{E}(x)$ has a weak $y$ dependence, we vary the fraction $f_y=1, 0.75, 0.5$ when computing the energy gain. The reported error bar for $g$ is the total spread due to these variations. 
The data analysis code and default parameters are available at \cite{Shi25code}.

\begin{figure*}[t]
	\centering
	\includegraphics[width=1.0\textwidth]{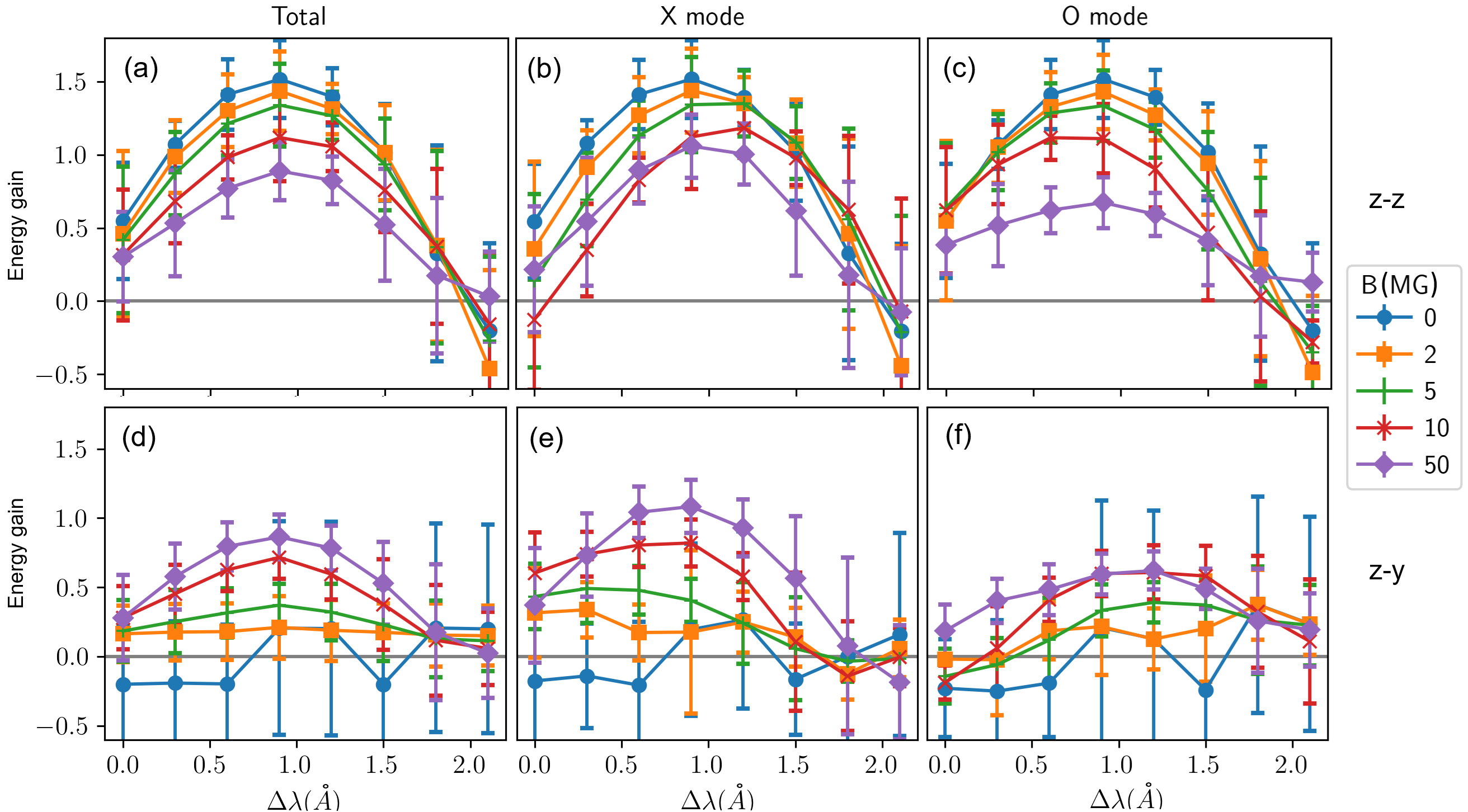} 
	\caption{Energy gain $g(\Delta\lambda)$ for MagCBET at $\theta_B=90^\circ$ and intermediate angle $\phi_B=30^\circ$. (a)-(c) For $z$-$z$ polarization, the gain reduces with $B$, and the IAW peak broadens and shifts. (d)-(f) For $z$-$y$ polarization, the gain increases with $B$. The total gain (a),(d) is within the X-mode gain (b), (e) and O-mode gain (c),(f). The gain at other $\theta_B$ and $\phi_B$ are qualitatively similar. }
	\label{fig:GainCurve}
\end{figure*}

\section{Results \label{sec:results}}
In this paper, we only show a representative subset of our data, and a more complete set of post processed PIC data is available at \cite{Shi25data}. 
First, we check our simulation setup and data analysis method using the $B=0$ case, which serves as a baseline for comparing magnetized cases. 
For simulations using the $z$-$y$ polarization setup, we verify that $g$ is nearly flat as a function of $\Delta\lambda$, which agrees with the expectation that orthogonally polarized lasers have zero CBET in unmagnetized plasmas. However, $g(\Delta\lambda)$ on average has a negative offset $g_0\approx-0.4$ due to losses from competing PIC processes, such as back and side scattering of the seed laser. 
To estimate how much competing processes depend on magnetization, we measure side scattered light that leaves from the $+y$ domain boundary. The scattered energy varies between $0.05\%$ and $0.17\%$ when $B$ changes from 0 to \mbox{20 MG}. Although this variation is large, the absolute amount is small. Because CBET is the dominant process, all gain values we report hereafter have the common offset $g_0$ removed. 
For the $z$-$z$ polarization, the gain curve shows a peak near the IAW resonances.
In the small-gain regime, we verify that $g(-\Delta \lambda)\approx - g(\Delta \lambda)$ is approximately an odd function, as expected from CBET theory. 
When the gain is larger, because of Eq.~(\ref{eq:gain_energy_amplitude}), the energy gain is no longer an odd function, but $g(\Delta\lambda=0)$ remains close to zero. 
We verify that when the pump inteisity $I_p$ is halved, $g$ is roughly halved for $\Delta \lambda>0$, because $G$ is proportional to the pump intensity. On the other hand, $g$ remains roughly unchanged for $\Delta\lambda<0$, because for negative $\Delta\lambda$, the roles of pump and seed lasers are swapped. 
Likewise, we verify that when the seed intensity $I_s$ is doubled, $g$ remains roughly unchanged for $\Delta\lambda>0$, whereas $g$ is roughly doubled for $\Delta\lambda<0$, consistent with CBET theory. After these basic checks, we will now focus on the positive half of the gain curve when discussing magnetized cases. 

\begin{figure}[t]
	\centering
	\includegraphics[width=1.0\linewidth]{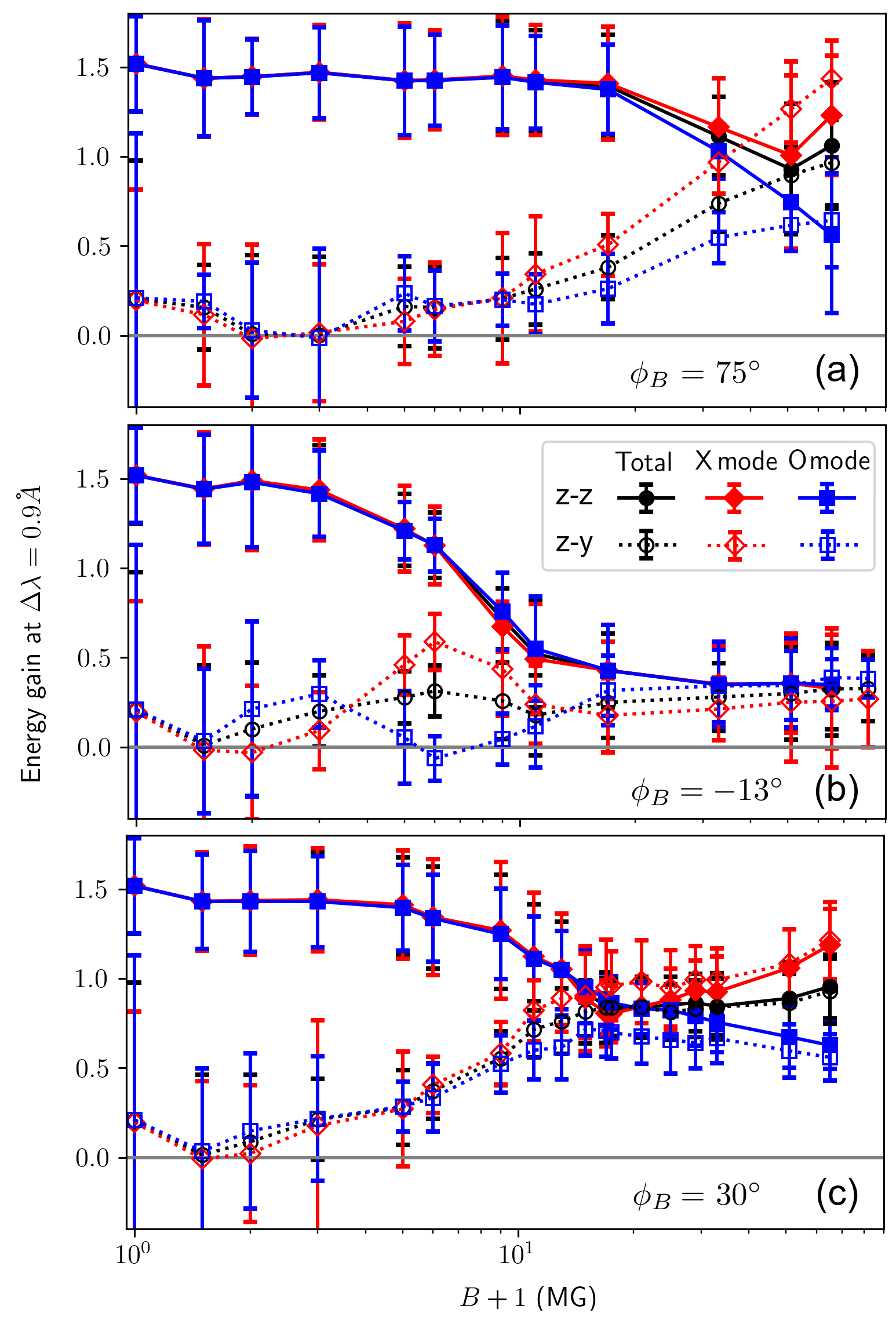} 
	\caption{Energy gain near the peak of the gain curves at $\theta_B=90^\circ$.
		(a) When the plasma wave $\Delta\mathbf{k}$ is near parallel to $\mathbf{B}$, the gains remain nearly constant until $B\sim$ \mbox{10 MG}, where laser modes become magnetized. 
		(b) When $\Delta\mathbf{k}$ is near perpendicular to $\mathbf{B}$, the plasma wave is magnetized. The gains are small at large $B$ and are insensitive to laser modes.
		(c) When $\Delta\mathbf{k}$ is at an intermediate angle to $\mathbf{B}$, both effects are at play, and the gains have features that resemble the two extreme cases. 
	}
	\label{fig:GainB}
\end{figure}

Second, we map out CBET gain curves for three magnetized cases for both polarizations. In all cases, we use $\theta_B=90^\circ$, which means that $\mathbf{B}$ is in the $x$-$y$ plane, namely, the crossing plane of the pump and seed lasers. 
(i) The case $\phi_B=75^\circ$ is when the plasma wavevector $\Delta\mathbf{k}=\mathbf{k}_0-\mathbf{k}$ is nearly parallel to the background magnetic field. In this case, the plasma wave that mediates CBET is essentially unmagnetized. Magnetization effects thus arise primarily from modifying the pump and seed lasers. The modifications remain small until $B\sim$ \mbox{10 MG}, when the electron cyclotron frequency 
is no longer negligible compared to the laser frequency. 
As $B$ increases, the gain for $z$-$z$ polarization reduces, while the gain for $z$-$y$ polarization increases. 
(ii) The case $\phi_B=-13^\circ$ is when the angle $\langle \Delta\mathbf{k}, \mathbf{B} \rangle\approx 88^\circ$ is near perpendicular. We avoid the exact perpendicular case, because it is a special case where some interactions may be exactly suppressed \cite{PhysRevE.96.023204, shi2018plasma, 10.1063/1.5099513}. In this near perpendicular case, the plasma wave is magnetized, and magnetization effects on CBET start to become decernible for $B\sim$ \mbox{1 MG}. When $B$ increases, the peak of the gain curve shifts and the gain curve generally becomes broader. While the gain for $z$-$z$ polarization reduces, for the $z$-$y$ polarization, the gain remains close to zero.
(iii) The case $\phi_B=30^\circ$ is at an intermediate angle $\langle \Delta\mathbf{k}, \mathbf{B} \rangle\approx 45^\circ$. The gain curves in this intermediate case are shown in Fig.~\ref{fig:GainCurve}, which have features that resemble the two extreme cases: Similar to the $\Delta\mathbf{k} \parallel \mathbf{B}$ case, the $z$-$z$ gain decreases, while the $z$-$y$ gain increases with $B$. Additionally, similar to the $\Delta\mathbf{k} \perp \mathbf{B}$ case, magnetization effects begin to manifest at a few MG, and the gain curves shift and broaden at higher $B$. 

Third, we scan the CBET gain as a function of the magnetic field strength at fixed $\Delta\lambda=0.9$ \AA, which is near the peak of the gain curves. 
The scans more clearly illustrate the dependence of $g$ on $B$. To show the unmagnetized case on a log scale, we plot $B+1$ MG for the abscissa, so the leftmost points corrspond to $B=0$. We again focus on the same three cases. 
(i) When $\Delta\mathbf{k}$ and $\mathbf{B}$ are near parallel ($\phi_B=75^\circ$), the gains remains roughly constant until $B\sim$ \mbox{10 MG}, as shown in Fig.~\ref{fig:GainB}(a). At weaker $B$, the X-mode gain (red diamonds) and O-mode gain (blue squares) are the same, and equal to the total gain (black circles). In other words, the pump laser transfers the same amount of energy to both eigenmodes, which are nearly degenerate in weak magnetic fields. At larger $B$, the two eigenmodes become non-degenerate, and their gains split from the total gain. Gains for the X mode, which is right-handed with respect to $\mathbf{B}$, is larger than gains for the O mode, which is left-handed with respect to $\mathbf{B}$. This is expected because the X mode couples more strongly with electron motion, which is also right-handed and has a larger response. 
The gains for the two polarizations meet at $B\sim$ \mbox{50 MG} and around half the unmagnetized $z$-$z$ gain. The crossing point depends on plasma conditions and the interaction length. 
(ii) When $\Delta\mathbf{k}$ and $\mathbf{B}$ are near perpendicular ($\phi_B=-13^\circ$), the gain for $z$-$z$ polarization decreases more rapidly with $B$, while the gain for $z$-$y$ polarization remains close to zero, as shown in Fig.~\ref{fig:GainB}(b). Beyond $B\sim$ \mbox{10 MG}, the gains drop below $\sim30\%$ of the unmagnetized gain. In this near perpendicular case, magnetization effects primarily arise from modifying the plasma wave, which provides a smaller coupling and is damped more heavily because cyclotron motion introduces additional phase mixing effects into the wave motion. The gain is insensitive to either X or O modes of the laser, whose gains are close to the total gain.  
Kinetic theories \cite{Boyd_Rankin_1985, Moody25} predict that at near perpendicular angles, additional resonances due to, for example, Bernstein waves also arise. We will report PIC simulations of these none-IAW mediated CBET in a separate paper. 
(iii) When $\Delta\mathbf{k}$ and $\mathbf{B}$ are at an intermediate angle ($\phi_B=30^\circ$), magnetization effects arise both from modifying the lasers and the mediating plasma wave, and the gains have an intermediate behavior, as shown in Fig.~\ref{fig:GainB}(c). Similar to the perpendicular case, magnetization effects begin to manifest at a few MG. 
Moreover, similar to the parallel case, the $z$-$z$ gain decreases, while the $z$-$y$ gain increases with $B$. The crossing point is at a smaller $B$ compared to the parallel case. The X-mode gain is larger than the O-mode gain, which sandwich the total gain.

Finally, we also perform simulations at different plasma conditions and interaction geometries, and find qualitatively similar behaviors. The additional simulations are for fully ionized hydrogen plasmas at density $n_e=3\times 10^{20}\;\text{cm}^{-3}$. The CBET crossing angle is $\alpha=24^\circ$. The field angles are $\theta_B=45^\circ$ and $\phi_B=15^\circ$, so that $\mathbf{B}$ has a component out of the crossing plane, and is at a general angle with respect to all waves participating in CBET.  
We map out gain curves at two temperatures. The gain at a lower temperature $T_e=4T_i=$ \mbox{0.4 keV} is overal larger than the gain at a higher temperature $T_e=4T_i=$ \mbox{1.2 keV}, similar to the unmagnetized case. Near the peak of the gain curves, the $z$-$z$ gain decreases while the $z$-$y$ gain increases with increasing $B$. 
At the lower temperature, the peak unmagnetized $z$-$z$ gain is $g\approx 2.1$, and $B= 5$ \mbox{MG} already noticeably suppress the gain to a peak value of $g\approx 1.8$. The $z$-$z$ and $z$-$y$ gains cross near $B\approx 18$ \mbox{MG}. At $B= 30$ \mbox{MG}, the gains are $\approx30\%$ of the unmagnetized $z$-$z$ gain. 
In comparison, at the higher temperature, a larger $B=$ \mbox{10 MG} is needed for noticeably suppresing the already-small gain, where the peak unmagnetized gain $g\approx 0.8$ is suppressed to $g\approx 0.6$. These additional data, not displayed in the main text, can be found at \cite{Shi25data}.

\section{Discussion\label{sec:discussion}}
We perform 2D-3V PIC simulations of MagCBET at fixed plasma density and temperature, which isolate direct magnetization effects on CBET. Starting from \mbox{$\sim$ MG fields}, direct magnetization effects arise from modifying the mediating IAW quasimode when it propagates at an angle with $\mathbf{B}$. Additionally, starting from \mbox{$\sim10$ MG} fields, direct magnetization effects arise from modifying the electromagnetic eigenmodes, which underly initially linearly polarized lasers that are commonly used in experiments. When two eigenmodes become nondegenerate, they mediate CBET at different rates, and the overall CBET is a superposition of eigenmode interactions. 
In experiments, when an external magnetic field is turned on, plasma conditions often become different. Appreciable indirect magnetization effects on CBET start at a lower \mbox{$\sim 10^{-1}$ MG} fields, and we will report our experimental findings in a forthcoming paper.

In this paper, we choose simulation parameters such that IAW mediated CBET is the dominant process and stays within the linear regime where pump depletion and distribution function evolution are insignificant. Outside this regime, CBET coexists with other LPI processes and the interactions are more complicated. The linear regime allows for a direct comparison with theory, which we will report in more details in our companion paper. On a qualitative level, our PIC simulations agree with the theory, which predicts three general trends when increasing the magnetic field strength.
First, when pump and seed laser polarizations are initially parallel, the gain decreases and the gain curve broadens and shifts. 
Second, when the polarizations are initially perpendicular, in which case unmagnetized CBET is forbidden, we find the gain increases beyond zero. The magnetic field at which gains of the two polarizations cross depends on plasma conditions and interaction geometry. 
Third, at a given polarization, X-mode gain is often larger than the O-mode gain, which split from the total gain. Overall, magnetization has a tendency to suppress CBET.

\textbf{Code and data availability.} 
The MATLAB code \cite{Shi25code} for performing data analysis is openly available on GitLab at \url{https://gitlab.com/seanYuanSHI/magnetized-cross-beam-energy-transfer},
where example EPOCH input deck and Linux shell script for setting up batch simulations can also be found. 
The data underlying Figs.~\ref{fig:GainCurve}-\ref{fig:GainB} are openly available on Zenodo at \url{https://zenodo.org/records/16498564}, where additional data that is discussed but not shown in this paper can also be found \cite{Shi25data}.

\begin{acknowledgments} 
	The authors thank Drs. David J. Strozzi, Luis S. Leal, and Bradley B. Pollock for helpful discussions. 
	This work utilized the Alpine high performance computing resource at the University of Colorado Boulder. Alpine is jointly funded by the University of Colorado Boulder, the University of Colorado Anschutz, Colorado State University, and the National Science Foundation (award 2201538).
	The EPOCH code was funded by the UK EPSRC grants EP/G054950/1, EP/G056803/1, EP/G055165/1 and EP/ M022463/1.
	This work is performed under the auspices of the U.S. Department of Energy by Lawrence
	Livermore National Laboratory under Contract DE-AC52-07NA27344 and by the LLNL-LDRD program under Project Number 23-ERD-025. 
\end{acknowledgments}


\begin{thebibliography}{94}%
	\makeatletter
	\providecommand \@ifxundefined [1]{%
		\@ifx{#1\undefined}
	}%
	\providecommand \@ifnum [1]{%
		\ifnum #1\expandafter \@firstoftwo
		\else \expandafter \@secondoftwo
		\fi
	}%
	\providecommand \@ifx [1]{%
		\ifx #1\expandafter \@firstoftwo
		\else \expandafter \@secondoftwo
		\fi
	}%
	\providecommand \natexlab [1]{#1}%
	\providecommand \enquote  [1]{``#1''}%
	\providecommand \bibnamefont  [1]{#1}%
	\providecommand \bibfnamefont [1]{#1}%
	\providecommand \citenamefont [1]{#1}%
	\providecommand \href@noop [0]{\@secondoftwo}%
	\providecommand \href [0]{\begingroup \@sanitize@url \@href}%
	\providecommand \@href[1]{\@@startlink{#1}\@@href}%
	\providecommand \@@href[1]{\endgroup#1\@@endlink}%
	\providecommand \@sanitize@url [0]{\catcode `\\12\catcode `\$12\catcode
		`\&12\catcode `\#12\catcode `\^12\catcode `\_12\catcode `\%12\relax}%
	\providecommand \@@startlink[1]{}%
	\providecommand \@@endlink[0]{}%
	\providecommand \url  [0]{\begingroup\@sanitize@url \@url }%
	\providecommand \@url [1]{\endgroup\@href {#1}{\urlprefix }}%
	\providecommand \urlprefix  [0]{URL }%
	\providecommand \Eprint [0]{\href }%
	\providecommand \doibase [0]{https://doi.org/}%
	\providecommand \selectlanguage [0]{\@gobble}%
	\providecommand \bibinfo  [0]{\@secondoftwo}%
	\providecommand \bibfield  [0]{\@secondoftwo}%
	\providecommand \translation [1]{[#1]}%
	\providecommand \BibitemOpen [0]{}%
	\providecommand \bibitemStop [0]{}%
	\providecommand \bibitemNoStop [0]{.\EOS\space}%
	\providecommand \EOS [0]{\spacefactor3000\relax}%
	\providecommand \BibitemShut  [1]{\csname bibitem#1\endcsname}%
	\let\auto@bib@innerbib\@empty
	\bibitem [{\citenamefont {Chang}\ \emph {et~al.}(2011)\citenamefont {Chang},
		\citenamefont {Fiksel}, \citenamefont {Hohenberger}, \citenamefont {Knauer},
		\citenamefont {Betti}, \citenamefont {Marshall}, \citenamefont {Meyerhofer},
		\citenamefont {S\'eguin},\ and\ \citenamefont {Petrasso}}]{Chang11}%
	\BibitemOpen
	\bibfield  {author} {\bibinfo {author} {\bibfnamefont {P.~Y.}\ \bibnamefont
			{Chang}}, \bibinfo {author} {\bibfnamefont {G.}~\bibnamefont {Fiksel}},
		\bibinfo {author} {\bibfnamefont {M.}~\bibnamefont {Hohenberger}}, \bibinfo
		{author} {\bibfnamefont {J.~P.}\ \bibnamefont {Knauer}}, \bibinfo {author}
		{\bibfnamefont {R.}~\bibnamefont {Betti}}, \bibinfo {author} {\bibfnamefont
			{F.~J.}\ \bibnamefont {Marshall}}, \bibinfo {author} {\bibfnamefont {D.~D.}\
			\bibnamefont {Meyerhofer}}, \bibinfo {author} {\bibfnamefont {F.~H.}\
			\bibnamefont {S\'eguin}},\ and\ \bibinfo {author} {\bibfnamefont {R.~D.}\
			\bibnamefont {Petrasso}},\ }\bibfield  {title} {\bibinfo {title} {Fusion
			yield enhancement in magnetized laser-driven implosions},\ }\href
	{https://doi.org/10.1103/PhysRevLett.107.035006} {\bibfield  {journal}
		{\bibinfo  {journal} {Physical Review Letters}\ }\textbf {\bibinfo {volume}
			{107}},\ \bibinfo {pages} {035006} (\bibinfo {year} {2011})}\BibitemShut
	{NoStop}%
	\bibitem [{\citenamefont {Hohenberger}\ \emph {et~al.}(2012)\citenamefont
		{Hohenberger}, \citenamefont {Chang}, \citenamefont {Fiksel}, \citenamefont
		{Knauer}, \citenamefont {Betti}, \citenamefont {Marshall}, \citenamefont
		{Meyerhofer}, \citenamefont {Séguin},\ and\ \citenamefont
		{Petrasso}}]{Hohenberger12}%
	\BibitemOpen
	\bibfield  {author} {\bibinfo {author} {\bibfnamefont {M.}~\bibnamefont
			{Hohenberger}}, \bibinfo {author} {\bibfnamefont {P.-Y.}\ \bibnamefont
			{Chang}}, \bibinfo {author} {\bibfnamefont {G.}~\bibnamefont {Fiksel}},
		\bibinfo {author} {\bibfnamefont {J.~P.}\ \bibnamefont {Knauer}}, \bibinfo
		{author} {\bibfnamefont {R.}~\bibnamefont {Betti}}, \bibinfo {author}
		{\bibfnamefont {F.~J.}\ \bibnamefont {Marshall}}, \bibinfo {author}
		{\bibfnamefont {D.~D.}\ \bibnamefont {Meyerhofer}}, \bibinfo {author}
		{\bibfnamefont {F.~H.}\ \bibnamefont {Séguin}},\ and\ \bibinfo {author}
		{\bibfnamefont {R.~D.}\ \bibnamefont {Petrasso}},\ }\bibfield  {title}
	{\bibinfo {title} {Inertial confinement fusion implosions with imposed
			magnetic field compression using the {OMEGA} lasera},\ }\href
	{https://doi.org/10.1063/1.3696032} {\bibfield  {journal} {\bibinfo
			{journal} {Physics of Plasmas}\ }\textbf {\bibinfo {volume} {19}},\ \bibinfo
		{pages} {056306} (\bibinfo {year} {2012})}\BibitemShut {NoStop}%
	\bibitem [{\citenamefont {Hansen}\ \emph {et~al.}(2020)\citenamefont {Hansen},
		\citenamefont {Davies}, \citenamefont {Barnak}, \citenamefont {Betti},
		\citenamefont {Campbell}, \citenamefont {Glebov}, \citenamefont {Knauer},
		\citenamefont {Leal}, \citenamefont {Peebles}, \citenamefont {Sefkow},\ and\
		\citenamefont {Woo}}]{Hansen20}%
	\BibitemOpen
	\bibfield  {author} {\bibinfo {author} {\bibfnamefont {E.~C.}\ \bibnamefont
			{Hansen}}, \bibinfo {author} {\bibfnamefont {J.~R.}\ \bibnamefont {Davies}},
		\bibinfo {author} {\bibfnamefont {D.~H.}\ \bibnamefont {Barnak}}, \bibinfo
		{author} {\bibfnamefont {R.}~\bibnamefont {Betti}}, \bibinfo {author}
		{\bibfnamefont {E.~M.}\ \bibnamefont {Campbell}}, \bibinfo {author}
		{\bibfnamefont {V.~Y.}\ \bibnamefont {Glebov}}, \bibinfo {author}
		{\bibfnamefont {J.~P.}\ \bibnamefont {Knauer}}, \bibinfo {author}
		{\bibfnamefont {L.~S.}\ \bibnamefont {Leal}}, \bibinfo {author}
		{\bibfnamefont {J.~L.}\ \bibnamefont {Peebles}}, \bibinfo {author}
		{\bibfnamefont {A.~B.}\ \bibnamefont {Sefkow}},\ and\ \bibinfo {author}
		{\bibfnamefont {K.~M.}\ \bibnamefont {Woo}},\ }\bibfield  {title} {\bibinfo
		{title} {Neutron yield enhancement and suppression by magnetization in
			laser-driven cylindrical implosions},\ }\href
	{https://doi.org/10.1063/1.5144447} {\bibfield  {journal} {\bibinfo
			{journal} {Physics of Plasmas}\ }\textbf {\bibinfo {volume} {27}},\ \bibinfo
		{pages} {062703} (\bibinfo {year} {2020})}\BibitemShut {NoStop}%
	\bibitem [{\citenamefont {Moody}\ \emph {et~al.}(2022)\citenamefont {Moody},
		\citenamefont {Pollock}, \citenamefont {Sio}, \citenamefont {Strozzi},
		\citenamefont {Ho}, \citenamefont {Walsh}, \citenamefont {Kemp},
		\citenamefont {Lahmann}, \citenamefont {Kucheyev}, \citenamefont
		{Kozioziemski}, \citenamefont {Carroll}, \citenamefont {Kroll}, \citenamefont
		{Yanagisawa}, \citenamefont {Angus}, \citenamefont {Bachmann}, \citenamefont
		{Bhandarkar}, \citenamefont {Bude}, \citenamefont {Divol}, \citenamefont
		{Ferguson}, \citenamefont {Fry}, \citenamefont {Hagler}, \citenamefont
		{Hartouni}, \citenamefont {Herrmann}, \citenamefont {Hsing}, \citenamefont
		{Holunga}, \citenamefont {Izumi}, \citenamefont {Javedani}, \citenamefont
		{Johnson}, \citenamefont {Khan}, \citenamefont {Kalantar}, \citenamefont
		{Kohut}, \citenamefont {Logan}, \citenamefont {Masters}, \citenamefont
		{Nikroo}, \citenamefont {Orsi}, \citenamefont {Piston}, \citenamefont
		{Provencher}, \citenamefont {Rowe}, \citenamefont {Sater}, \citenamefont
		{Skulina}, \citenamefont {Stygar}, \citenamefont {Tang}, \citenamefont
		{Winters}, \citenamefont {Zimmerman}, \citenamefont {Adrian}, \citenamefont
		{Chittenden}, \citenamefont {Appelbe}, \citenamefont {Boxall}, \citenamefont
		{Crilly}, \citenamefont {O'Neill}, \citenamefont {Davies}, \citenamefont
		{Peebles},\ and\ \citenamefont {Fujioka}}]{Moody22}%
	\BibitemOpen
	\bibfield  {author} {\bibinfo {author} {\bibfnamefont {J.~D.}\ \bibnamefont
			{Moody}}, \bibinfo {author} {\bibfnamefont {B.~B.}\ \bibnamefont {Pollock}},
		\bibinfo {author} {\bibfnamefont {H.}~\bibnamefont {Sio}}, \bibinfo {author}
		{\bibfnamefont {D.~J.}\ \bibnamefont {Strozzi}}, \bibinfo {author}
		{\bibfnamefont {D.~D.-M.}\ \bibnamefont {Ho}}, \bibinfo {author}
		{\bibfnamefont {C.~A.}\ \bibnamefont {Walsh}}, \bibinfo {author}
		{\bibfnamefont {G.~E.}\ \bibnamefont {Kemp}}, \bibinfo {author}
		{\bibfnamefont {B.}~\bibnamefont {Lahmann}}, \bibinfo {author} {\bibfnamefont
			{S.~O.}\ \bibnamefont {Kucheyev}}, \bibinfo {author} {\bibfnamefont
			{B.}~\bibnamefont {Kozioziemski}}, \bibinfo {author} {\bibfnamefont {E.~G.}\
			\bibnamefont {Carroll}}, \bibinfo {author} {\bibfnamefont {J.}~\bibnamefont
			{Kroll}}, \bibinfo {author} {\bibfnamefont {D.~K.}\ \bibnamefont
			{Yanagisawa}}, \bibinfo {author} {\bibfnamefont {J.}~\bibnamefont {Angus}},
		\bibinfo {author} {\bibfnamefont {B.}~\bibnamefont {Bachmann}}, \bibinfo
		{author} {\bibfnamefont {S.~D.}\ \bibnamefont {Bhandarkar}}, \bibinfo
		{author} {\bibfnamefont {J.~D.}\ \bibnamefont {Bude}}, \bibinfo {author}
		{\bibfnamefont {L.}~\bibnamefont {Divol}}, \bibinfo {author} {\bibfnamefont
			{B.}~\bibnamefont {Ferguson}}, \bibinfo {author} {\bibfnamefont
			{J.}~\bibnamefont {Fry}}, \bibinfo {author} {\bibfnamefont {L.}~\bibnamefont
			{Hagler}}, \bibinfo {author} {\bibfnamefont {E.}~\bibnamefont {Hartouni}},
		\bibinfo {author} {\bibfnamefont {M.~C.}\ \bibnamefont {Herrmann}}, \bibinfo
		{author} {\bibfnamefont {W.}~\bibnamefont {Hsing}}, \bibinfo {author}
		{\bibfnamefont {D.~M.}\ \bibnamefont {Holunga}}, \bibinfo {author}
		{\bibfnamefont {N.}~\bibnamefont {Izumi}}, \bibinfo {author} {\bibfnamefont
			{J.}~\bibnamefont {Javedani}}, \bibinfo {author} {\bibfnamefont
			{A.}~\bibnamefont {Johnson}}, \bibinfo {author} {\bibfnamefont
			{S.}~\bibnamefont {Khan}}, \bibinfo {author} {\bibfnamefont {D.}~\bibnamefont
			{Kalantar}}, \bibinfo {author} {\bibfnamefont {T.}~\bibnamefont {Kohut}},
		\bibinfo {author} {\bibfnamefont {B.~G.}\ \bibnamefont {Logan}}, \bibinfo
		{author} {\bibfnamefont {N.}~\bibnamefont {Masters}}, \bibinfo {author}
		{\bibfnamefont {A.}~\bibnamefont {Nikroo}}, \bibinfo {author} {\bibfnamefont
			{N.}~\bibnamefont {Orsi}}, \bibinfo {author} {\bibfnamefont {K.}~\bibnamefont
			{Piston}}, \bibinfo {author} {\bibfnamefont {C.}~\bibnamefont {Provencher}},
		\bibinfo {author} {\bibfnamefont {A.}~\bibnamefont {Rowe}}, \bibinfo {author}
		{\bibfnamefont {J.}~\bibnamefont {Sater}}, \bibinfo {author} {\bibfnamefont
			{K.}~\bibnamefont {Skulina}}, \bibinfo {author} {\bibfnamefont {W.~A.}\
			\bibnamefont {Stygar}}, \bibinfo {author} {\bibfnamefont {V.}~\bibnamefont
			{Tang}}, \bibinfo {author} {\bibfnamefont {S.~E.}\ \bibnamefont {Winters}},
		\bibinfo {author} {\bibfnamefont {G.}~\bibnamefont {Zimmerman}}, \bibinfo
		{author} {\bibfnamefont {P.}~\bibnamefont {Adrian}}, \bibinfo {author}
		{\bibfnamefont {J.~P.}\ \bibnamefont {Chittenden}}, \bibinfo {author}
		{\bibfnamefont {B.}~\bibnamefont {Appelbe}}, \bibinfo {author} {\bibfnamefont
			{A.}~\bibnamefont {Boxall}}, \bibinfo {author} {\bibfnamefont
			{A.}~\bibnamefont {Crilly}}, \bibinfo {author} {\bibfnamefont
			{S.}~\bibnamefont {O'Neill}}, \bibinfo {author} {\bibfnamefont
			{J.}~\bibnamefont {Davies}}, \bibinfo {author} {\bibfnamefont
			{J.}~\bibnamefont {Peebles}},\ and\ \bibinfo {author} {\bibfnamefont
			{S.}~\bibnamefont {Fujioka}},\ }\bibfield  {title} {\bibinfo {title}
		{Increased ion temperature and neutron yield observed in magnetized
			indirectly driven ${\mathrm{d}}_{2}$-filled capsule implosions on the
			{National Ignition Facility}},\ }\href
	{https://doi.org/10.1103/PhysRevLett.129.195002} {\bibfield  {journal}
		{\bibinfo  {journal} {Physical Review Letters}\ }\textbf {\bibinfo {volume}
			{129}},\ \bibinfo {pages} {195002} (\bibinfo {year} {2022})}\BibitemShut
	{NoStop}%
	\bibitem [{\citenamefont {Peebles}\ \emph {et~al.}(2023)\citenamefont
		{Peebles}, \citenamefont {Davies}, \citenamefont {Barnak}, \citenamefont
		{Glebov}, \citenamefont {Hansen}, \citenamefont {Heuer}, \citenamefont
		{Leal}, \citenamefont {Bonino}, \citenamefont {Harding}, \citenamefont
		{Sefkow}, \citenamefont {Peterson}, \citenamefont {Sinars}, \citenamefont
		{Campbell},\ and\ \citenamefont {Betti}}]{Peebles23}%
	\BibitemOpen
	\bibfield  {author} {\bibinfo {author} {\bibfnamefont {J.~L.}\ \bibnamefont
			{Peebles}}, \bibinfo {author} {\bibfnamefont {J.~R.}\ \bibnamefont {Davies}},
		\bibinfo {author} {\bibfnamefont {D.~H.}\ \bibnamefont {Barnak}}, \bibinfo
		{author} {\bibfnamefont {V.~Y.}\ \bibnamefont {Glebov}}, \bibinfo {author}
		{\bibfnamefont {E.~C.}\ \bibnamefont {Hansen}}, \bibinfo {author}
		{\bibfnamefont {P.~V.}\ \bibnamefont {Heuer}}, \bibinfo {author}
		{\bibfnamefont {L.~S.}\ \bibnamefont {Leal}}, \bibinfo {author}
		{\bibfnamefont {M.~J.}\ \bibnamefont {Bonino}}, \bibinfo {author}
		{\bibfnamefont {D.~R.}\ \bibnamefont {Harding}}, \bibinfo {author}
		{\bibfnamefont {A.~B.}\ \bibnamefont {Sefkow}}, \bibinfo {author}
		{\bibfnamefont {K.~J.}\ \bibnamefont {Peterson}}, \bibinfo {author}
		{\bibfnamefont {D.~B.}\ \bibnamefont {Sinars}}, \bibinfo {author}
		{\bibfnamefont {E.~M.}\ \bibnamefont {Campbell}},\ and\ \bibinfo {author}
		{\bibfnamefont {R.}~\bibnamefont {Betti}},\ }\bibfield  {title} {\bibinfo
		{title} {Demonstration of neutron-yield enhancement by laser preheating and
			magnetization of laser-driven cylindrical implosions},\ }\href
	{https://doi.org/10.1063/5.0159653} {\bibfield  {journal} {\bibinfo
			{journal} {Physics of Plasmas}\ }\textbf {\bibinfo {volume} {30}},\ \bibinfo
		{pages} {082703} (\bibinfo {year} {2023})}\BibitemShut {NoStop}%
	\bibitem [{\citenamefont {Perkins}\ \emph {et~al.}(2017)\citenamefont
		{Perkins}, \citenamefont {Ho}, \citenamefont {Logan}, \citenamefont
		{Zimmerman}, \citenamefont {Rhodes}, \citenamefont {Strozzi}, \citenamefont
		{Blackfield},\ and\ \citenamefont {Hawkins}}]{Perkins17}%
	\BibitemOpen
	\bibfield  {author} {\bibinfo {author} {\bibfnamefont {L.~J.}\ \bibnamefont
			{Perkins}}, \bibinfo {author} {\bibfnamefont {D.~D.-M.}\ \bibnamefont {Ho}},
		\bibinfo {author} {\bibfnamefont {B.~G.}\ \bibnamefont {Logan}}, \bibinfo
		{author} {\bibfnamefont {G.~B.}\ \bibnamefont {Zimmerman}}, \bibinfo {author}
		{\bibfnamefont {M.~A.}\ \bibnamefont {Rhodes}}, \bibinfo {author}
		{\bibfnamefont {D.~J.}\ \bibnamefont {Strozzi}}, \bibinfo {author}
		{\bibfnamefont {D.~T.}\ \bibnamefont {Blackfield}},\ and\ \bibinfo {author}
		{\bibfnamefont {S.~A.}\ \bibnamefont {Hawkins}},\ }\bibfield  {title}
	{\bibinfo {title} {The potential of imposed magnetic fields for enhancing
			ignition probability and fusion energy yield in indirect-drive inertial
			confinement fusion},\ }\href {https://doi.org/10.1063/1.4985150} {\bibfield
		{journal} {\bibinfo  {journal} {Physics of Plasmas}\ }\textbf {\bibinfo
			{volume} {24}},\ \bibinfo {pages} {062708} (\bibinfo {year}
		{2017})}\BibitemShut {NoStop}%
	\bibitem [{\citenamefont {Walsh}\ \emph {et~al.}(2022)\citenamefont {Walsh},
		\citenamefont {O'Neill}, \citenamefont {Chittenden}, \citenamefont {Crilly},
		\citenamefont {Appelbe}, \citenamefont {Strozzi}, \citenamefont {Ho},
		\citenamefont {Sio}, \citenamefont {Pollock}, \citenamefont {Divol},
		\citenamefont {Hartouni}, \citenamefont {Rosen}, \citenamefont {Logan},\ and\
		\citenamefont {Moody}}]{Walsh22}%
	\BibitemOpen
	\bibfield  {author} {\bibinfo {author} {\bibfnamefont {C.~A.}\ \bibnamefont
			{Walsh}}, \bibinfo {author} {\bibfnamefont {S.}~\bibnamefont {O'Neill}},
		\bibinfo {author} {\bibfnamefont {J.~P.}\ \bibnamefont {Chittenden}},
		\bibinfo {author} {\bibfnamefont {A.~J.}\ \bibnamefont {Crilly}}, \bibinfo
		{author} {\bibfnamefont {B.}~\bibnamefont {Appelbe}}, \bibinfo {author}
		{\bibfnamefont {D.~J.}\ \bibnamefont {Strozzi}}, \bibinfo {author}
		{\bibfnamefont {D.}~\bibnamefont {Ho}}, \bibinfo {author} {\bibfnamefont
			{H.}~\bibnamefont {Sio}}, \bibinfo {author} {\bibfnamefont {B.}~\bibnamefont
			{Pollock}}, \bibinfo {author} {\bibfnamefont {L.}~\bibnamefont {Divol}},
		\bibinfo {author} {\bibfnamefont {E.}~\bibnamefont {Hartouni}}, \bibinfo
		{author} {\bibfnamefont {M.}~\bibnamefont {Rosen}}, \bibinfo {author}
		{\bibfnamefont {B.~G.}\ \bibnamefont {Logan}},\ and\ \bibinfo {author}
		{\bibfnamefont {J.~D.}\ \bibnamefont {Moody}},\ }\bibfield  {title} {\bibinfo
		{title} {Magnetized {ICF} implosions: Scaling of temperature and yield
			enhancement},\ }\href {https://doi.org/10.1063/5.0081915} {\bibfield
		{journal} {\bibinfo  {journal} {Physics of Plasmas}\ }\textbf {\bibinfo
			{volume} {29}},\ \bibinfo {pages} {042701} (\bibinfo {year}
		{2022})}\BibitemShut {NoStop}%
	\bibitem [{\citenamefont {Strozzi}\ \emph {et~al.}(2024)\citenamefont
		{Strozzi}, \citenamefont {Sio}, \citenamefont {Zimmerman}, \citenamefont
		{Moody}, \citenamefont {Weber}, \citenamefont {Djordjević}, \citenamefont
		{Walsh}, \citenamefont {Hammel}, \citenamefont {Pollock}, \citenamefont
		{Povilus}, \citenamefont {Chittenden},\ and\ \citenamefont
		{O'Neill}}]{Strozzi24}%
	\BibitemOpen
	\bibfield  {author} {\bibinfo {author} {\bibfnamefont {D.~J.}\ \bibnamefont
			{Strozzi}}, \bibinfo {author} {\bibfnamefont {H.}~\bibnamefont {Sio}},
		\bibinfo {author} {\bibfnamefont {G.~B.}\ \bibnamefont {Zimmerman}}, \bibinfo
		{author} {\bibfnamefont {J.~D.}\ \bibnamefont {Moody}}, \bibinfo {author}
		{\bibfnamefont {C.~R.}\ \bibnamefont {Weber}}, \bibinfo {author}
		{\bibfnamefont {B.~Z.}\ \bibnamefont {Djordjević}}, \bibinfo {author}
		{\bibfnamefont {C.~A.}\ \bibnamefont {Walsh}}, \bibinfo {author}
		{\bibfnamefont {B.~A.}\ \bibnamefont {Hammel}}, \bibinfo {author}
		{\bibfnamefont {B.~B.}\ \bibnamefont {Pollock}}, \bibinfo {author}
		{\bibfnamefont {A.}~\bibnamefont {Povilus}}, \bibinfo {author} {\bibfnamefont
			{J.~P.}\ \bibnamefont {Chittenden}},\ and\ \bibinfo {author} {\bibfnamefont
			{S.}~\bibnamefont {O'Neill}},\ }\bibfield  {title} {\bibinfo {title} {Design
			and modeling of indirectly driven magnetized implosions on the {NIF}},\
	}\href {https://doi.org/10.1063/5.0214674} {\bibfield  {journal} {\bibinfo
			{journal} {Physics of Plasmas}\ }\textbf {\bibinfo {volume} {31}},\ \bibinfo
		{pages} {092703} (\bibinfo {year} {2024})}\BibitemShut {NoStop}%
	\bibitem [{\citenamefont {Walsh}\ \emph {et~al.}(2025)\citenamefont {Walsh},
		\citenamefont {O'Neill}, \citenamefont {Strozzi}, \citenamefont {Leal},
		\citenamefont {Spiers}, \citenamefont {Crilly}, \citenamefont {Pollock},
		\citenamefont {Sio}, \citenamefont {Hammel}, \citenamefont {Djordjević},
		\citenamefont {Hurricane}, \citenamefont {Chittenden},\ and\ \citenamefont
		{Moody}}]{Walsh25}%
	\BibitemOpen
	\bibfield  {author} {\bibinfo {author} {\bibfnamefont {C.~A.}\ \bibnamefont
			{Walsh}}, \bibinfo {author} {\bibfnamefont {S.~T.}\ \bibnamefont {O'Neill}},
		\bibinfo {author} {\bibfnamefont {D.~J.}\ \bibnamefont {Strozzi}}, \bibinfo
		{author} {\bibfnamefont {L.~S.}\ \bibnamefont {Leal}}, \bibinfo {author}
		{\bibfnamefont {R.}~\bibnamefont {Spiers}}, \bibinfo {author} {\bibfnamefont
			{A.~J.}\ \bibnamefont {Crilly}}, \bibinfo {author} {\bibfnamefont
			{B.}~\bibnamefont {Pollock}}, \bibinfo {author} {\bibfnamefont
			{H.}~\bibnamefont {Sio}}, \bibinfo {author} {\bibfnamefont {B.}~\bibnamefont
			{Hammel}}, \bibinfo {author} {\bibfnamefont {B.~Z.}\ \bibnamefont
			{Djordjević}}, \bibinfo {author} {\bibfnamefont {O.}~\bibnamefont
			{Hurricane}}, \bibinfo {author} {\bibfnamefont {J.~P.}\ \bibnamefont
			{Chittenden}},\ and\ \bibinfo {author} {\bibfnamefont {J.~D.}\ \bibnamefont
			{Moody}},\ }\bibfield  {title} {\bibinfo {title} {Magnetized {ICF}
			implosions: Ignition at low laser energy using designs with more ablator mass
			remaining},\ }\href {https://doi.org/10.1063/5.0274275} {\bibfield  {journal}
		{\bibinfo  {journal} {Physics of Plasmas}\ }\textbf {\bibinfo {volume}
			{32}},\ \bibinfo {pages} {072715} (\bibinfo {year} {2025})}\BibitemShut
	{NoStop}%
	\bibitem [{\citenamefont {Appelbe}\ \emph {et~al.}(2021)\citenamefont
		{Appelbe}, \citenamefont {Velikovich}, \citenamefont {Sherlock},
		\citenamefont {Walsh}, \citenamefont {Crilly}, \citenamefont {O'~Neill},\
		and\ \citenamefont {Chittenden}}]{Appelbe21}%
	\BibitemOpen
	\bibfield  {author} {\bibinfo {author} {\bibfnamefont {B.}~\bibnamefont
			{Appelbe}}, \bibinfo {author} {\bibfnamefont {A.~L.}\ \bibnamefont
			{Velikovich}}, \bibinfo {author} {\bibfnamefont {M.}~\bibnamefont
			{Sherlock}}, \bibinfo {author} {\bibfnamefont {C.}~\bibnamefont {Walsh}},
		\bibinfo {author} {\bibfnamefont {A.}~\bibnamefont {Crilly}}, \bibinfo
		{author} {\bibfnamefont {S.}~\bibnamefont {O'~Neill}},\ and\ \bibinfo
		{author} {\bibfnamefont {J.}~\bibnamefont {Chittenden}},\ }\bibfield  {title}
	{\bibinfo {title} {Magnetic field transport in propagating thermonuclear
			burn},\ }\href {https://doi.org/10.1063/5.0040161} {\bibfield  {journal}
		{\bibinfo  {journal} {Physics of Plasmas}\ }\textbf {\bibinfo {volume}
			{28}},\ \bibinfo {pages} {032705} (\bibinfo {year} {2021})}\BibitemShut
	{NoStop}%
	\bibitem [{\citenamefont {O'Neill}\ \emph {et~al.}(2025)\citenamefont
		{O'Neill}, \citenamefont {Appelbe}, \citenamefont {Crilly}, \citenamefont
		{Walsh}, \citenamefont {Strozzi}, \citenamefont {Moody},\ and\ \citenamefont
		{Chittenden}}]{Neill25}%
	\BibitemOpen
	\bibfield  {author} {\bibinfo {author} {\bibfnamefont {S.~T.}\ \bibnamefont
			{O'Neill}}, \bibinfo {author} {\bibfnamefont {B.~D.}\ \bibnamefont
			{Appelbe}}, \bibinfo {author} {\bibfnamefont {A.~J.}\ \bibnamefont {Crilly}},
		\bibinfo {author} {\bibfnamefont {C.~A.}\ \bibnamefont {Walsh}}, \bibinfo
		{author} {\bibfnamefont {D.~J.}\ \bibnamefont {Strozzi}}, \bibinfo {author}
		{\bibfnamefont {J.~D.}\ \bibnamefont {Moody}},\ and\ \bibinfo {author}
		{\bibfnamefont {J.~P.}\ \bibnamefont {Chittenden}},\ }\bibfield  {title}
	{\bibinfo {title} {Burn propagation in magnetized high-yield inertial
			fusion},\ }\href {https://doi.org/10.1063/5.0242215} {\bibfield  {journal}
		{\bibinfo  {journal} {Physics of Plasmas}\ }\textbf {\bibinfo {volume}
			{32}},\ \bibinfo {pages} {022703} (\bibinfo {year} {2025})}\BibitemShut
	{NoStop}%
	\bibitem [{\citenamefont {Stamper}(1991)}]{Stamper91}%
	\BibitemOpen
	\bibfield  {author} {\bibinfo {author} {\bibfnamefont {J.~A.}\ \bibnamefont
			{Stamper}},\ }\bibfield  {title} {\bibinfo {title} {Review on spontaneous
			magnetic fields in laser-produced plasmas: Phenomena and measurements},\
	}\href {https://doi.org/10.1017/S0263034600006595} {\bibfield  {journal}
		{\bibinfo  {journal} {Laser and Particle Beams}\ }\textbf {\bibinfo {volume}
			{9}},\ \bibinfo {pages} {841–862} (\bibinfo {year} {1991})}\BibitemShut
	{NoStop}%
	\bibitem [{\citenamefont {Sherlock}\ and\ \citenamefont
		{Bissell}(2020)}]{Sherlock20}%
	\BibitemOpen
	\bibfield  {author} {\bibinfo {author} {\bibfnamefont {M.}~\bibnamefont
			{Sherlock}}\ and\ \bibinfo {author} {\bibfnamefont {J.~J.}\ \bibnamefont
			{Bissell}},\ }\bibfield  {title} {\bibinfo {title} {Suppression of the
			{Biermann} battery and stabilization of the thermomagnetic instability in
			laser fusion conditions},\ }\href
	{https://doi.org/10.1103/PhysRevLett.124.055001} {\bibfield  {journal}
		{\bibinfo  {journal} {Physical Review Letters}\ }\textbf {\bibinfo {volume}
			{124}},\ \bibinfo {pages} {055001} (\bibinfo {year} {2020})}\BibitemShut
	{NoStop}%
	\bibitem [{\citenamefont {Walsh}\ \emph {et~al.}(2021)\citenamefont {Walsh},
		\citenamefont {Sadler},\ and\ \citenamefont {Davies}}]{Walsh21}%
	\BibitemOpen
	\bibfield  {author} {\bibinfo {author} {\bibfnamefont {C.}~\bibnamefont
			{Walsh}}, \bibinfo {author} {\bibfnamefont {J.}~\bibnamefont {Sadler}},\ and\
		\bibinfo {author} {\bibfnamefont {J.}~\bibnamefont {Davies}},\ }\bibfield
	{title} {\bibinfo {title} {Updated magnetized transport coefficients: impact
			on laser-plasmas with self-generated or applied magnetic fields},\ }\href
	{https://doi.org/10.1088/1741-4326/ac25c1} {\bibfield  {journal} {\bibinfo
			{journal} {Nuclear Fusion}\ }\textbf {\bibinfo {volume} {61}},\ \bibinfo
		{pages} {116025} (\bibinfo {year} {2021})}\BibitemShut {NoStop}%
	\bibitem [{\citenamefont {Campbell}\ \emph {et~al.}(2022)\citenamefont
		{Campbell}, \citenamefont {Walsh}, \citenamefont {Russell}, \citenamefont
		{Chittenden}, \citenamefont {Crilly}, \citenamefont {Fiksel}, \citenamefont
		{Gao}, \citenamefont {Igumenshchev}, \citenamefont {Nilson}, \citenamefont
		{Thomas}, \citenamefont {Krushelnick},\ and\ \citenamefont
		{Willingale}}]{Campbell22}%
	\BibitemOpen
	\bibfield  {author} {\bibinfo {author} {\bibfnamefont {P.~T.}\ \bibnamefont
			{Campbell}}, \bibinfo {author} {\bibfnamefont {C.~A.}\ \bibnamefont {Walsh}},
		\bibinfo {author} {\bibfnamefont {B.~K.}\ \bibnamefont {Russell}}, \bibinfo
		{author} {\bibfnamefont {J.~P.}\ \bibnamefont {Chittenden}}, \bibinfo
		{author} {\bibfnamefont {A.}~\bibnamefont {Crilly}}, \bibinfo {author}
		{\bibfnamefont {G.}~\bibnamefont {Fiksel}}, \bibinfo {author} {\bibfnamefont
			{L.}~\bibnamefont {Gao}}, \bibinfo {author} {\bibfnamefont {I.~V.}\
			\bibnamefont {Igumenshchev}}, \bibinfo {author} {\bibfnamefont {P.~M.}\
			\bibnamefont {Nilson}}, \bibinfo {author} {\bibfnamefont {A.~G.~R.}\
			\bibnamefont {Thomas}}, \bibinfo {author} {\bibfnamefont {K.}~\bibnamefont
			{Krushelnick}},\ and\ \bibinfo {author} {\bibfnamefont {L.}~\bibnamefont
			{Willingale}},\ }\bibfield  {title} {\bibinfo {title} {Measuring magnetic
			flux suppression in high-power laser–plasma interactions},\ }\href
	{https://doi.org/10.1063/5.0062717} {\bibfield  {journal} {\bibinfo
			{journal} {Physics of Plasmas}\ }\textbf {\bibinfo {volume} {29}},\ \bibinfo
		{pages} {012701} (\bibinfo {year} {2022})}\BibitemShut {NoStop}%
	\bibitem [{\citenamefont {Cattani}\ and\ \citenamefont
		{Sacchi}(1966)}]{Cattani66}%
	\BibitemOpen
	\bibfield  {author} {\bibinfo {author} {\bibfnamefont {D.}~\bibnamefont
			{Cattani}}\ and\ \bibinfo {author} {\bibfnamefont {C.}~\bibnamefont
			{Sacchi}},\ }\bibfield  {title} {\bibinfo {title} {A theory on the creation
			of stellar magnetic fields},\ }\href@noop {} {\bibfield  {journal} {\bibinfo
			{journal} {Il Nuovo Cimento B (1965-1970)}\ }\textbf {\bibinfo {volume}
			{46}},\ \bibinfo {pages} {258} (\bibinfo {year} {1966})}\BibitemShut
	{NoStop}%
	\bibitem [{\citenamefont {Munirov}\ and\ \citenamefont
		{Fisch}(2017)}]{Munirov17}%
	\BibitemOpen
	\bibfield  {author} {\bibinfo {author} {\bibfnamefont {V.~R.}\ \bibnamefont
			{Munirov}}\ and\ \bibinfo {author} {\bibfnamefont {N.~J.}\ \bibnamefont
			{Fisch}},\ }\bibfield  {title} {\bibinfo {title} {Radiative transfer dynamo
			effect},\ }\href {https://doi.org/10.1103/PhysRevE.95.013205} {\bibfield
		{journal} {\bibinfo  {journal} {Physical Review E}\ }\textbf {\bibinfo
			{volume} {95}},\ \bibinfo {pages} {013205} (\bibinfo {year}
		{2017})}\BibitemShut {NoStop}%
	\bibitem [{\citenamefont {Liu}\ and\ \citenamefont {Zhang}(2024)}]{Liu24}%
	\BibitemOpen
	\bibfield  {author} {\bibinfo {author} {\bibfnamefont {Y.}~\bibnamefont
			{Liu}}\ and\ \bibinfo {author} {\bibfnamefont {G.-F.}\ \bibnamefont
			{Zhang}},\ }\bibfield  {title} {\bibinfo {title} {Self-generated magnetic
			field in laser plasma with super-{Gaussian} distributed electrons},\ }\href
	{https://doi.org/10.1063/5.0229891} {\bibfield  {journal} {\bibinfo
			{journal} {Physics of Plasmas}\ }\textbf {\bibinfo {volume} {31}},\ \bibinfo
		{pages} {092303} (\bibinfo {year} {2024})}\BibitemShut {NoStop}%
	\bibitem [{\citenamefont {Gotchev}\ \emph {et~al.}(2009)\citenamefont
		{Gotchev}, \citenamefont {Chang}, \citenamefont {Knauer}, \citenamefont
		{Meyerhofer}, \citenamefont {Polomarov}, \citenamefont {Frenje},
		\citenamefont {Li}, \citenamefont {Manuel}, \citenamefont {Petrasso},
		\citenamefont {Rygg}, \citenamefont {S\'eguin},\ and\ \citenamefont
		{Betti}}]{Gotchev09}%
	\BibitemOpen
	\bibfield  {author} {\bibinfo {author} {\bibfnamefont {O.~V.}\ \bibnamefont
			{Gotchev}}, \bibinfo {author} {\bibfnamefont {P.~Y.}\ \bibnamefont {Chang}},
		\bibinfo {author} {\bibfnamefont {J.~P.}\ \bibnamefont {Knauer}}, \bibinfo
		{author} {\bibfnamefont {D.~D.}\ \bibnamefont {Meyerhofer}}, \bibinfo
		{author} {\bibfnamefont {O.}~\bibnamefont {Polomarov}}, \bibinfo {author}
		{\bibfnamefont {J.}~\bibnamefont {Frenje}}, \bibinfo {author} {\bibfnamefont
			{C.~K.}\ \bibnamefont {Li}}, \bibinfo {author} {\bibfnamefont {M.~J.-E.}\
			\bibnamefont {Manuel}}, \bibinfo {author} {\bibfnamefont {R.~D.}\
			\bibnamefont {Petrasso}}, \bibinfo {author} {\bibfnamefont {J.~R.}\
			\bibnamefont {Rygg}}, \bibinfo {author} {\bibfnamefont {F.~H.}\ \bibnamefont
			{S\'eguin}},\ and\ \bibinfo {author} {\bibfnamefont {R.}~\bibnamefont
			{Betti}},\ }\bibfield  {title} {\bibinfo {title} {Laser-driven magnetic-flux
			compression in high-energy-density plasmas},\ }\href
	{https://doi.org/10.1103/PhysRevLett.103.215004} {\bibfield  {journal}
		{\bibinfo  {journal} {Physical Review Letters}\ }\textbf {\bibinfo {volume}
			{103}},\ \bibinfo {pages} {215004} (\bibinfo {year} {2009})}\BibitemShut
	{NoStop}%
	\bibitem [{\citenamefont {Knauer}\ \emph {et~al.}(2010)\citenamefont {Knauer},
		\citenamefont {Gotchev}, \citenamefont {Chang}, \citenamefont {Meyerhofer},
		\citenamefont {Polomarov}, \citenamefont {Betti}, \citenamefont {Frenje},
		\citenamefont {Li}, \citenamefont {Manuel}, \citenamefont {Petrasso},
		\citenamefont {Rygg},\ and\ \citenamefont {S\'guin}}]{Knauer10}%
	\BibitemOpen
	\bibfield  {author} {\bibinfo {author} {\bibfnamefont {J.~P.}\ \bibnamefont
			{Knauer}}, \bibinfo {author} {\bibfnamefont {O.~V.}\ \bibnamefont {Gotchev}},
		\bibinfo {author} {\bibfnamefont {P.~Y.}\ \bibnamefont {Chang}}, \bibinfo
		{author} {\bibfnamefont {D.~D.}\ \bibnamefont {Meyerhofer}}, \bibinfo
		{author} {\bibfnamefont {O.}~\bibnamefont {Polomarov}}, \bibinfo {author}
		{\bibfnamefont {R.}~\bibnamefont {Betti}}, \bibinfo {author} {\bibfnamefont
			{J.~A.}\ \bibnamefont {Frenje}}, \bibinfo {author} {\bibfnamefont {C.~K.}\
			\bibnamefont {Li}}, \bibinfo {author} {\bibfnamefont {M.~J.-E.}\ \bibnamefont
			{Manuel}}, \bibinfo {author} {\bibfnamefont {R.~D.}\ \bibnamefont
			{Petrasso}}, \bibinfo {author} {\bibfnamefont {J.~R.}\ \bibnamefont {Rygg}},\
		and\ \bibinfo {author} {\bibfnamefont {F.~H.}\ \bibnamefont {S\'guin}},\
	}\bibfield  {title} {\bibinfo {title} {Compressing magnetic fields with
			high-energy lasers},\ }\href {https://doi.org/10.1063/1.3416557} {\bibfield
		{journal} {\bibinfo  {journal} {Physics of Plasmas}\ }\textbf {\bibinfo
			{volume} {17}},\ \bibinfo {pages} {056318} (\bibinfo {year}
		{2010})}\BibitemShut {NoStop}%
	\bibitem [{\citenamefont {Farmer}\ \emph {et~al.}(2017)\citenamefont {Farmer},
		\citenamefont {Koning}, \citenamefont {Strozzi}, \citenamefont {Hinkel},
		\citenamefont {Berzak~Hopkins}, \citenamefont {Jones},\ and\ \citenamefont
		{Rosen}}]{Farmer17}%
	\BibitemOpen
	\bibfield  {author} {\bibinfo {author} {\bibfnamefont {W.~A.}\ \bibnamefont
			{Farmer}}, \bibinfo {author} {\bibfnamefont {J.~M.}\ \bibnamefont {Koning}},
		\bibinfo {author} {\bibfnamefont {D.~J.}\ \bibnamefont {Strozzi}}, \bibinfo
		{author} {\bibfnamefont {D.~E.}\ \bibnamefont {Hinkel}}, \bibinfo {author}
		{\bibfnamefont {L.~F.}\ \bibnamefont {Berzak~Hopkins}}, \bibinfo {author}
		{\bibfnamefont {O.~S.}\ \bibnamefont {Jones}},\ and\ \bibinfo {author}
		{\bibfnamefont {M.~D.}\ \bibnamefont {Rosen}},\ }\bibfield  {title} {\bibinfo
		{title} {Simulation of self-generated magnetic fields in an inertial fusion
			hohlraum environment},\ }\href {https://doi.org/10.1063/1.4983140} {\bibfield
		{journal} {\bibinfo  {journal} {Physics of Plasmas}\ }\textbf {\bibinfo
			{volume} {24}},\ \bibinfo {pages} {052703} (\bibinfo {year}
		{2017})}\BibitemShut {NoStop}%
	\bibitem [{\citenamefont {Sio}\ \emph {et~al.}(2021)\citenamefont {Sio},
		\citenamefont {Moody}, \citenamefont {Ho}, \citenamefont {Pollock},
		\citenamefont {Walsh}, \citenamefont {Lahmann}, \citenamefont {Strozzi},
		\citenamefont {Kemp}, \citenamefont {Hsing}, \citenamefont {Crilly},
		\citenamefont {Chittenden},\ and\ \citenamefont {Appelbe}}]{Sio21}%
	\BibitemOpen
	\bibfield  {author} {\bibinfo {author} {\bibfnamefont {H.}~\bibnamefont
			{Sio}}, \bibinfo {author} {\bibfnamefont {J.~D.}\ \bibnamefont {Moody}},
		\bibinfo {author} {\bibfnamefont {D.~D.}\ \bibnamefont {Ho}}, \bibinfo
		{author} {\bibfnamefont {B.~B.}\ \bibnamefont {Pollock}}, \bibinfo {author}
		{\bibfnamefont {C.~A.}\ \bibnamefont {Walsh}}, \bibinfo {author}
		{\bibfnamefont {B.}~\bibnamefont {Lahmann}}, \bibinfo {author} {\bibfnamefont
			{D.~J.}\ \bibnamefont {Strozzi}}, \bibinfo {author} {\bibfnamefont {G.~E.}\
			\bibnamefont {Kemp}}, \bibinfo {author} {\bibfnamefont {W.~W.}\ \bibnamefont
			{Hsing}}, \bibinfo {author} {\bibfnamefont {A.}~\bibnamefont {Crilly}},
		\bibinfo {author} {\bibfnamefont {J.~P.}\ \bibnamefont {Chittenden}},\ and\
		\bibinfo {author} {\bibfnamefont {B.}~\bibnamefont {Appelbe}},\ }\bibfield
	{title} {\bibinfo {title} {Diagnosing plasma magnetization in inertial
			confinement fusion implosions using secondary deuterium-tritium reactions},\
	}\href {https://doi.org/10.1063/5.0043381} {\bibfield  {journal} {\bibinfo
			{journal} {Review of Scientific Instruments}\ }\textbf {\bibinfo {volume}
			{92}},\ \bibinfo {pages} {043543} (\bibinfo {year} {2021})}\BibitemShut
	{NoStop}%
	\bibitem [{\citenamefont {Shi}\ \emph {et~al.}(2018)\citenamefont {Shi},
		\citenamefont {Qin},\ and\ \citenamefont {Fisch}}]{Shi18a}%
	\BibitemOpen
	\bibfield  {author} {\bibinfo {author} {\bibfnamefont {Y.}~\bibnamefont
			{Shi}}, \bibinfo {author} {\bibfnamefont {H.}~\bibnamefont {Qin}},\ and\
		\bibinfo {author} {\bibfnamefont {N.~J.}\ \bibnamefont {Fisch}},\ }\bibfield
	{title} {\bibinfo {title} {Laser-plasma interactions in magnetized
			environment},\ }\href {https://doi.org/10.1063/1.5017980} {\bibfield
		{journal} {\bibinfo  {journal} {Physics of Plasmas}\ }\textbf {\bibinfo
			{volume} {25}},\ \bibinfo {pages} {055706} (\bibinfo {year}
		{2018})}\BibitemShut {NoStop}%
	\bibitem [{\citenamefont {Edwards}\ \emph {et~al.}(2019)\citenamefont
		{Edwards}, \citenamefont {Shi}, \citenamefont {Mikhailova},\ and\
		\citenamefont {Fisch}}]{Edwards19}%
	\BibitemOpen
	\bibfield  {author} {\bibinfo {author} {\bibfnamefont {M.~R.}\ \bibnamefont
			{Edwards}}, \bibinfo {author} {\bibfnamefont {Y.}~\bibnamefont {Shi}},
		\bibinfo {author} {\bibfnamefont {J.~M.}\ \bibnamefont {Mikhailova}},\ and\
		\bibinfo {author} {\bibfnamefont {N.~J.}\ \bibnamefont {Fisch}},\ }\bibfield
	{title} {\bibinfo {title} {Laser amplification in strongly magnetized
			plasma},\ }\href {https://doi.org/10.1103/PhysRevLett.123.025001} {\bibfield
		{journal} {\bibinfo  {journal} {Physical Review Letters}\ }\textbf {\bibinfo
			{volume} {123}},\ \bibinfo {pages} {025001} (\bibinfo {year}
		{2019})}\BibitemShut {NoStop}%
	\bibitem [{\citenamefont {Manzo}\ \emph {et~al.}(2022)\citenamefont {Manzo},
		\citenamefont {Edwards},\ and\ \citenamefont {Shi}}]{Manzo22}%
	\BibitemOpen
	\bibfield  {author} {\bibinfo {author} {\bibfnamefont {L.}~\bibnamefont
			{Manzo}}, \bibinfo {author} {\bibfnamefont {M.~R.}\ \bibnamefont {Edwards}},\
		and\ \bibinfo {author} {\bibfnamefont {Y.}~\bibnamefont {Shi}},\ }\bibfield
	{title} {\bibinfo {title} {Enhanced collisionless laser absorption in
			strongly magnetized plasmas},\ }\href {https://doi.org/10.1063/5.0100727}
	{\bibfield  {journal} {\bibinfo  {journal} {Physics of Plasmas}\ }\textbf
		{\bibinfo {volume} {29}},\ \bibinfo {pages} {112704} (\bibinfo {year}
		{2022})}\BibitemShut {NoStop}%
	\bibitem [{\citenamefont {Shi}(2018)}]{shi2018plasma}%
	\BibitemOpen
	\bibfield  {author} {\bibinfo {author} {\bibfnamefont {Y.}~\bibnamefont
			{Shi}},\ }\emph {\bibinfo {title} {Plasma physics in strong-field regimes}},\
	\href@noop {} {Ph.D. thesis},\ \bibinfo  {school} {Princeton University}
	(\bibinfo {year} {2018})\BibitemShut {NoStop}%
	\bibitem [{\citenamefont {Shi}\ \emph {et~al.}(2021)\citenamefont {Shi},
		\citenamefont {Qin},\ and\ \citenamefont {Fisch}}]{Shi21}%
	\BibitemOpen
	\bibfield  {author} {\bibinfo {author} {\bibfnamefont {Y.}~\bibnamefont
			{Shi}}, \bibinfo {author} {\bibfnamefont {H.}~\bibnamefont {Qin}},\ and\
		\bibinfo {author} {\bibfnamefont {N.~J.}\ \bibnamefont {Fisch}},\ }\bibfield
	{title} {\bibinfo {title} {Plasma physics in strong-field regimes: Theories
			and simulations},\ }\href {https://doi.org/10.1063/5.0043228} {\bibfield
		{journal} {\bibinfo  {journal} {Physics of Plasmas}\ }\textbf {\bibinfo
			{volume} {28}},\ \bibinfo {pages} {042104} (\bibinfo {year}
		{2021})}\BibitemShut {NoStop}%
	\bibitem [{\citenamefont {Shi}(2023)}]{Shi23}%
	\BibitemOpen
	\bibfield  {author} {\bibinfo {author} {\bibfnamefont {Y.}~\bibnamefont
			{Shi}},\ }\bibfield  {title} {\bibinfo {title} {Benchmarking magnetised
			three-wave coupling for laser backscattering: analytic solutions and kinetic
			simulations},\ }\href {https://doi.org/10.1017/S0022377823000405} {\bibfield
		{journal} {\bibinfo  {journal} {Journal of Plasma Physics}\ }\textbf
		{\bibinfo {volume} {89}},\ \bibinfo {pages} {905890305} (\bibinfo {year}
		{2023})}\BibitemShut {NoStop}%
	\bibitem [{\citenamefont {Boyd}\ and\ \citenamefont
		{Rankin}(1985)}]{Boyd_Rankin_1985}%
	\BibitemOpen
	\bibfield  {author} {\bibinfo {author} {\bibfnamefont {T.~J.~M.}\
			\bibnamefont {Boyd}}\ and\ \bibinfo {author} {\bibfnamefont {R.}~\bibnamefont
			{Rankin}},\ }\bibfield  {title} {\bibinfo {title} {Kinetic theory of
			stimulated {Raman} scattering from a magnetized plasma},\ }\href
	{https://doi.org/10.1017/S002237780000252X} {\bibfield  {journal} {\bibinfo
			{journal} {Journal of Plasma Physics}\ }\textbf {\bibinfo {volume} {33}},\
		\bibinfo {pages} {303–319} (\bibinfo {year} {1985})}\BibitemShut {NoStop}%
	\bibitem [{\citenamefont {Yao}\ \emph {et~al.}(2023)\citenamefont {Yao},
		\citenamefont {Higginson}, \citenamefont {Marqu\`es}, \citenamefont {Antici},
		\citenamefont {B\'eard}, \citenamefont {Burdonov}, \citenamefont {Borghesi},
		\citenamefont {Castan}, \citenamefont {Ciardi}, \citenamefont {Coleman},
		\citenamefont {Chen}, \citenamefont {d'Humi\`eres}, \citenamefont {Gangolf},
		\citenamefont {Gremillet}, \citenamefont {Khiar}, \citenamefont {Lancia},
		\citenamefont {Loiseau}, \citenamefont {Ribeyre}, \citenamefont {Soloviev},
		\citenamefont {Starodubtsev}, \citenamefont {Wang},\ and\ \citenamefont
		{Fuchs}}]{Yao23}%
	\BibitemOpen
	\bibfield  {author} {\bibinfo {author} {\bibfnamefont {W.}~\bibnamefont
			{Yao}}, \bibinfo {author} {\bibfnamefont {A.}~\bibnamefont {Higginson}},
		\bibinfo {author} {\bibfnamefont {J.-R.}\ \bibnamefont {Marqu\`es}}, \bibinfo
		{author} {\bibfnamefont {P.}~\bibnamefont {Antici}}, \bibinfo {author}
		{\bibfnamefont {J.}~\bibnamefont {B\'eard}}, \bibinfo {author} {\bibfnamefont
			{K.}~\bibnamefont {Burdonov}}, \bibinfo {author} {\bibfnamefont
			{M.}~\bibnamefont {Borghesi}}, \bibinfo {author} {\bibfnamefont
			{A.}~\bibnamefont {Castan}}, \bibinfo {author} {\bibfnamefont
			{A.}~\bibnamefont {Ciardi}}, \bibinfo {author} {\bibfnamefont
			{B.}~\bibnamefont {Coleman}}, \bibinfo {author} {\bibfnamefont {S.~N.}\
			\bibnamefont {Chen}}, \bibinfo {author} {\bibfnamefont {E.}~\bibnamefont
			{d'Humi\`eres}}, \bibinfo {author} {\bibfnamefont {T.}~\bibnamefont
			{Gangolf}}, \bibinfo {author} {\bibfnamefont {L.}~\bibnamefont {Gremillet}},
		\bibinfo {author} {\bibfnamefont {B.}~\bibnamefont {Khiar}}, \bibinfo
		{author} {\bibfnamefont {L.}~\bibnamefont {Lancia}}, \bibinfo {author}
		{\bibfnamefont {P.}~\bibnamefont {Loiseau}}, \bibinfo {author} {\bibfnamefont
			{X.}~\bibnamefont {Ribeyre}}, \bibinfo {author} {\bibfnamefont
			{A.}~\bibnamefont {Soloviev}}, \bibinfo {author} {\bibfnamefont
			{M.}~\bibnamefont {Starodubtsev}}, \bibinfo {author} {\bibfnamefont
			{Q.}~\bibnamefont {Wang}},\ and\ \bibinfo {author} {\bibfnamefont
			{J.}~\bibnamefont {Fuchs}},\ }\bibfield  {title} {\bibinfo {title} {Dynamics
			of nanosecond laser pulse propagation and of associated instabilities in a
			magnetized underdense plasma},\ }\href
	{https://doi.org/10.1103/PhysRevLett.130.265101} {\bibfield  {journal}
		{\bibinfo  {journal} {Physical Review Letters}\ }\textbf {\bibinfo {volume}
			{130}},\ \bibinfo {pages} {265101} (\bibinfo {year} {2023})}\BibitemShut
	{NoStop}%
	\bibitem [{\citenamefont {Winjum}\ \emph {et~al.}(2018)\citenamefont {Winjum},
		\citenamefont {Tsung},\ and\ \citenamefont {Mori}}]{Winjum18}%
	\BibitemOpen
	\bibfield  {author} {\bibinfo {author} {\bibfnamefont {B.~J.}\ \bibnamefont
			{Winjum}}, \bibinfo {author} {\bibfnamefont {F.~S.}\ \bibnamefont {Tsung}},\
		and\ \bibinfo {author} {\bibfnamefont {W.~B.}\ \bibnamefont {Mori}},\
	}\bibfield  {title} {\bibinfo {title} {Mitigation of stimulated {Raman}
			scattering in the kinetic regime by external magnetic fields},\ }\href
	{https://doi.org/10.1103/PhysRevE.98.043208} {\bibfield  {journal} {\bibinfo
			{journal} {Physical Review E}\ }\textbf {\bibinfo {volume} {98}},\ \bibinfo
		{pages} {043208} (\bibinfo {year} {2018})}\BibitemShut {NoStop}%
	\bibitem [{\citenamefont {Laham}\ \emph {et~al.}(1998)\citenamefont {Laham},
		\citenamefont {Nasser},\ and\ \citenamefont {Khateeb}}]{Nabil98}%
	\BibitemOpen
	\bibfield  {author} {\bibinfo {author} {\bibfnamefont {N.~M.}\ \bibnamefont
			{Laham}}, \bibinfo {author} {\bibfnamefont {A.~S.~A.}\ \bibnamefont
			{Nasser}},\ and\ \bibinfo {author} {\bibfnamefont {A.~M.}\ \bibnamefont
			{Khateeb}},\ }\bibfield  {title} {\bibinfo {title} {Effects of axial magnetic
			fields on backward {Raman} scattering in inhomogeneous plasmas},\ }\href
	{https://doi.org/10.1088/0031-8949/57/2/018} {\bibfield  {journal} {\bibinfo
			{journal} {Physica Scripta}\ }\textbf {\bibinfo {volume} {57}},\ \bibinfo
		{pages} {253} (\bibinfo {year} {1998})}\BibitemShut {NoStop}%
	\bibitem [{\citenamefont {Galloway}\ and\ \citenamefont
		{Kim}(1971)}]{Galloway_Kim_1971}%
	\BibitemOpen
	\bibfield  {author} {\bibinfo {author} {\bibfnamefont {J.~J.}\ \bibnamefont
			{Galloway}}\ and\ \bibinfo {author} {\bibfnamefont {H.}~\bibnamefont {Kim}},\
	}\bibfield  {title} {\bibinfo {title} {Lagrangian approach to non-linear wave
			interactions in a warm plasma},\ }\href
	{https://doi.org/10.1017/S002237780002568X} {\bibfield  {journal} {\bibinfo
			{journal} {Journal of Plasma Physics}\ }\textbf {\bibinfo {volume} {6}},\
		\bibinfo {pages} {53–72} (\bibinfo {year} {1971})}\BibitemShut {NoStop}%
	\bibitem [{\citenamefont {Boyd}\ and\ \citenamefont {Turner}(1978)}]{Boyd78}%
	\BibitemOpen
	\bibfield  {author} {\bibinfo {author} {\bibfnamefont {T.~J.~M.}\
			\bibnamefont {Boyd}}\ and\ \bibinfo {author} {\bibfnamefont {J.~G.}\
			\bibnamefont {Turner}},\ }\bibfield  {title} {\bibinfo {title} {Three‐ and
			four‐wave interactions in plasmas},\ }\href
	{https://doi.org/10.1063/1.523842} {\bibfield  {journal} {\bibinfo  {journal}
			{Journal of Mathematical Physics}\ }\textbf {\bibinfo {volume} {19}},\
		\bibinfo {pages} {1403} (\bibinfo {year} {1978})}\BibitemShut {NoStop}%
	\bibitem [{\citenamefont {Liu}\ and\ \citenamefont
		{Tripathi}(1986)}]{Liu1986143}%
	\BibitemOpen
	\bibfield  {author} {\bibinfo {author} {\bibfnamefont {C.}~\bibnamefont
			{Liu}}\ and\ \bibinfo {author} {\bibfnamefont {V.}~\bibnamefont {Tripathi}},\
	}\bibfield  {title} {\bibinfo {title} {Parametric instabilities in a
			magnetized plasma},\ }\href
	{https://doi.org/https://doi.org/10.1016/0370-1573(86)90108-0} {\bibfield
		{journal} {\bibinfo  {journal} {Physics Reports}\ }\textbf {\bibinfo {volume}
			{130}},\ \bibinfo {pages} {143} (\bibinfo {year} {1986})}\BibitemShut
	{NoStop}%
	\bibitem [{\citenamefont {Cohen}(1987)}]{Cohen87}%
	\BibitemOpen
	\bibfield  {author} {\bibinfo {author} {\bibfnamefont {B.~I.}\ \bibnamefont
			{Cohen}},\ }\bibfield  {title} {\bibinfo {title} {Compact dispersion
			relations for parametric instabilities of electromagnetic waves in magnetized
			plasmas},\ }\href {https://doi.org/10.1063/1.866032} {\bibfield  {journal}
		{\bibinfo  {journal} {The Physics of Fluids}\ }\textbf {\bibinfo {volume}
			{30}},\ \bibinfo {pages} {2676} (\bibinfo {year} {1987})}\BibitemShut
	{NoStop}%
	\bibitem [{\citenamefont {Porkolab}(1978)}]{Porkolab_1978}%
	\BibitemOpen
	\bibfield  {author} {\bibinfo {author} {\bibfnamefont {M.}~\bibnamefont
			{Porkolab}},\ }\bibfield  {title} {\bibinfo {title} {Parametric processes in
			magnetically confined ctr plasmas},\ }\href
	{https://doi.org/10.1088/0029-5515/18/3/008} {\bibfield  {journal} {\bibinfo
			{journal} {Nuclear Fusion}\ }\textbf {\bibinfo {volume} {18}},\ \bibinfo
		{pages} {367} (\bibinfo {year} {1978})}\BibitemShut {NoStop}%
	\bibitem [{\citenamefont {Grebogi}\ and\ \citenamefont
		{Liu}(1980)}]{Grebogi-Liu-80}%
	\BibitemOpen
	\bibfield  {author} {\bibinfo {author} {\bibfnamefont {C.}~\bibnamefont
			{Grebogi}}\ and\ \bibinfo {author} {\bibfnamefont {C.~S.}\ \bibnamefont
			{Liu}},\ }\bibfield  {title} {\bibinfo {title} {Brillouin and raman
			scattering of an extraordinary mode in a magnetized plasma},\ }\href
	{https://doi.org/10.1063/1.863146} {\bibfield  {journal} {\bibinfo  {journal}
			{The Physics of Fluids}\ }\textbf {\bibinfo {volume} {23}},\ \bibinfo {pages}
		{1330} (\bibinfo {year} {1980})}\BibitemShut {NoStop}%
	\bibitem [{\citenamefont {Hansen}\ \emph {et~al.}(2017)\citenamefont {Hansen},
		\citenamefont {Nielsen}, \citenamefont {Salewski}, \citenamefont {Stejner},
		\citenamefont {Stober},\ and\ \citenamefont {the ASDEX
			Upgrade~team}}]{Hansen_2017}%
	\BibitemOpen
	\bibfield  {author} {\bibinfo {author} {\bibfnamefont {S.~K.}\ \bibnamefont
			{Hansen}}, \bibinfo {author} {\bibfnamefont {S.~K.}\ \bibnamefont {Nielsen}},
		\bibinfo {author} {\bibfnamefont {M.}~\bibnamefont {Salewski}}, \bibinfo
		{author} {\bibfnamefont {M.}~\bibnamefont {Stejner}}, \bibinfo {author}
		{\bibfnamefont {J.}~\bibnamefont {Stober}},\ and\ \bibinfo {author}
		{\bibnamefont {the ASDEX Upgrade~team}},\ }\bibfield  {title} {\bibinfo
		{title} {Parametric decay instability near the upper hybrid resonance in
			magnetically confined fusion plasmas},\ }\href
	{https://doi.org/10.1088/1361-6587/aa7978} {\bibfield  {journal} {\bibinfo
			{journal} {Plasma Physics and Controlled Fusion}\ }\textbf {\bibinfo {volume}
			{59}},\ \bibinfo {pages} {105006} (\bibinfo {year} {2017})}\BibitemShut
	{NoStop}%
	\bibitem [{\citenamefont {Platzman}\ \emph {et~al.}(1968)\citenamefont
		{Platzman}, \citenamefont {Wolff},\ and\ \citenamefont {Tzoar}}]{Platzman68}%
	\BibitemOpen
	\bibfield  {author} {\bibinfo {author} {\bibfnamefont {P.~M.}\ \bibnamefont
			{Platzman}}, \bibinfo {author} {\bibfnamefont {P.~A.}\ \bibnamefont
			{Wolff}},\ and\ \bibinfo {author} {\bibfnamefont {N.}~\bibnamefont {Tzoar}},\
	}\bibfield  {title} {\bibinfo {title} {Light scattering from a plasma in a
			magnetic field},\ }\href {https://doi.org/10.1103/PhysRev.174.489} {\bibfield
		{journal} {\bibinfo  {journal} {Physical Reviews}\ }\textbf {\bibinfo
			{volume} {174}},\ \bibinfo {pages} {489} (\bibinfo {year}
		{1968})}\BibitemShut {NoStop}%
	\bibitem [{\citenamefont {Stenflo}(1981{\natexlab{a}})}]{Stenflo81}%
	\BibitemOpen
	\bibfield  {author} {\bibinfo {author} {\bibfnamefont {L.}~\bibnamefont
			{Stenflo}},\ }\bibfield  {title} {\bibinfo {title} {Comments on stimulated
			scattering of electromagnetic waves by electron bernstein modes in a
			plasma},\ }\href {https://doi.org/10.1103/PhysRevA.23.2730} {\bibfield
		{journal} {\bibinfo  {journal} {Physical Review A}\ }\textbf {\bibinfo
			{volume} {23}},\ \bibinfo {pages} {2730} (\bibinfo {year}
		{1981}{\natexlab{a}})}\BibitemShut {NoStop}%
	\bibitem [{\citenamefont {Kasymov}\ \emph {et~al.}(1985)\citenamefont
		{Kasymov}, \citenamefont {N{\"a}slund}, \citenamefont {Starodub},\ and\
		\citenamefont {Stenflo}}]{kasymov1985upper}%
	\BibitemOpen
	\bibfield  {author} {\bibinfo {author} {\bibfnamefont {Z.~Z.}\ \bibnamefont
			{Kasymov}}, \bibinfo {author} {\bibfnamefont {E.}~\bibnamefont
			{N{\"a}slund}}, \bibinfo {author} {\bibfnamefont {A.}~\bibnamefont
			{Starodub}},\ and\ \bibinfo {author} {\bibfnamefont {L.}~\bibnamefont
			{Stenflo}},\ }\bibfield  {title} {\bibinfo {title} {Upper hybrid turbulence
			in a plasma with magnetized electrons},\ }\href@noop {} {\bibfield  {journal}
		{\bibinfo  {journal} {Physica Scripta}\ }\textbf {\bibinfo {volume} {31}},\
		\bibinfo {pages} {201} (\bibinfo {year} {1985})}\BibitemShut {NoStop}%
	\bibitem [{\citenamefont {Hasegawa}\ and\ \citenamefont
		{Chen}(1976)}]{Hasegawa-Liu_76}%
	\BibitemOpen
	\bibfield  {author} {\bibinfo {author} {\bibfnamefont {A.}~\bibnamefont
			{Hasegawa}}\ and\ \bibinfo {author} {\bibfnamefont {L.}~\bibnamefont
			{Chen}},\ }\bibfield  {title} {\bibinfo {title} {Kinetic processes in plasma
			heating by resonant mode conversion of {Alfvén} wave},\ }\href
	{https://doi.org/10.1063/1.861427} {\bibfield  {journal} {\bibinfo  {journal}
			{The Physics of Fluids}\ }\textbf {\bibinfo {volume} {19}},\ \bibinfo {pages}
		{1924} (\bibinfo {year} {1976})}\BibitemShut {NoStop}%
	\bibitem [{\citenamefont {Wong}\ and\ \citenamefont
		{Goldstein}(1986)}]{Wong-Goldstein86}%
	\BibitemOpen
	\bibfield  {author} {\bibinfo {author} {\bibfnamefont {H.~K.}\ \bibnamefont
			{Wong}}\ and\ \bibinfo {author} {\bibfnamefont {M.~L.}\ \bibnamefont
			{Goldstein}},\ }\bibfield  {title} {\bibinfo {title} {Parametric
			instabilities of circularly polarized alfvén waves including dispersion},\
	}\href {https://doi.org/https://doi.org/10.1029/JA091iA05p05617} {\bibfield
		{journal} {\bibinfo  {journal} {Journal of Geophysical Research: Space
				Physics}\ }\textbf {\bibinfo {volume} {91}},\ \bibinfo {pages} {5617}
		(\bibinfo {year} {1986})}\BibitemShut {NoStop}%
	\bibitem [{\citenamefont {{Viñas}}\ and\ \citenamefont
		{Goldstein}(1991)}]{Viñas_Goldstein_1991}%
	\BibitemOpen
	\bibfield  {author} {\bibinfo {author} {\bibfnamefont {A.~F.}\ \bibnamefont
			{{Viñas}}}\ and\ \bibinfo {author} {\bibfnamefont {M.~L.}\ \bibnamefont
			{Goldstein}},\ }\bibfield  {title} {\bibinfo {title} {Parametric
			instabilities of circularly polarized large-amplitude dispersive {Alfvén}
			waves: excitation of parallel-propagating electromagnetic daughter waves},\
	}\href {https://doi.org/10.1017/S0022377800015981} {\bibfield  {journal}
		{\bibinfo  {journal} {Journal of Plasma Physics}\ }\textbf {\bibinfo {volume}
			{46}},\ \bibinfo {pages} {107–127} (\bibinfo {year} {1991})}\BibitemShut
	{NoStop}%
	\bibitem [{\citenamefont {Sj{\"o}lund}\ and\ \citenamefont
		{Stenflo}(1967{\natexlab{a}})}]{Sjolund67}%
	\BibitemOpen
	\bibfield  {author} {\bibinfo {author} {\bibfnamefont {A.}~\bibnamefont
			{Sj{\"o}lund}}\ and\ \bibinfo {author} {\bibfnamefont {L.}~\bibnamefont
			{Stenflo}},\ }\bibfield  {title} {\bibinfo {title} {Non-linear coupling in a
			magnetized plasma},\ }\href@noop {} {\bibfield  {journal} {\bibinfo
			{journal} {Zeitschrift für Physik A Hadrons and Nuclei}\ }\textbf {\bibinfo
			{volume} {204}},\ \bibinfo {pages} {211} (\bibinfo {year}
		{1967}{\natexlab{a}})}\BibitemShut {NoStop}%
	\bibitem [{\citenamefont {Sj{\"o}lund}\ and\ \citenamefont
		{Stenflo}(1967{\natexlab{b}})}]{sjolund1967parametric}%
	\BibitemOpen
	\bibfield  {author} {\bibinfo {author} {\bibfnamefont {A.}~\bibnamefont
			{Sj{\"o}lund}}\ and\ \bibinfo {author} {\bibfnamefont {L.}~\bibnamefont
			{Stenflo}},\ }\bibfield  {title} {\bibinfo {title} {Parametric coupling
			between transverse electromagnetic and longitudinal electron waves},\
	}\href@noop {} {\bibfield  {journal} {\bibinfo  {journal} {Physica}\ }\textbf
		{\bibinfo {volume} {35}},\ \bibinfo {pages} {499} (\bibinfo {year}
		{1967}{\natexlab{b}})}\BibitemShut {NoStop}%
	\bibitem [{\citenamefont {Stenflo}\ \emph {et~al.}(1970)\citenamefont
		{Stenflo}, \citenamefont {Weiland},\ and\ \citenamefont
		{Wilhelmsson}}]{stenflo1970solution}%
	\BibitemOpen
	\bibfield  {author} {\bibinfo {author} {\bibfnamefont {L.}~\bibnamefont
			{Stenflo}}, \bibinfo {author} {\bibfnamefont {J.}~\bibnamefont {Weiland}},\
		and\ \bibinfo {author} {\bibfnamefont {H.}~\bibnamefont {Wilhelmsson}},\
	}\bibfield  {title} {\bibinfo {title} {A solution of equations describing
			explosive instabilities},\ }\href@noop {} {\bibfield  {journal} {\bibinfo
			{journal} {Physica Scripta}\ }\textbf {\bibinfo {volume} {1}},\ \bibinfo
		{pages} {46} (\bibinfo {year} {1970})}\BibitemShut {NoStop}%
	\bibitem [{\citenamefont {Stenflo}\ \emph {et~al.}(1971)\citenamefont
		{Stenflo}, \citenamefont {Wilhelmsson},\ and\ \citenamefont
		{{\"O}stberg}}]{stenflo1971non}%
	\BibitemOpen
	\bibfield  {author} {\bibinfo {author} {\bibfnamefont {L.}~\bibnamefont
			{Stenflo}}, \bibinfo {author} {\bibfnamefont {H.}~\bibnamefont
			{Wilhelmsson}},\ and\ \bibinfo {author} {\bibfnamefont {K.}~\bibnamefont
			{{\"O}stberg}},\ }\bibfield  {title} {\bibinfo {title} {Non-linear
			instabilities in streaming plasmas},\ }\href@noop {} {\bibfield  {journal}
		{\bibinfo  {journal} {Physica Scripta}\ }\textbf {\bibinfo {volume} {3}},\
		\bibinfo {pages} {231} (\bibinfo {year} {1971})}\BibitemShut {NoStop}%
	\bibitem [{\citenamefont {Stenflo}(1973)}]{stenflo1973three}%
	\BibitemOpen
	\bibfield  {author} {\bibinfo {author} {\bibfnamefont {L.}~\bibnamefont
			{Stenflo}},\ }\bibfield  {title} {\bibinfo {title} {Three-wave interaction in
			cold magnetized plasmas},\ }\href@noop {} {\bibfield  {journal} {\bibinfo
			{journal} {Planetary and Space Science}\ }\textbf {\bibinfo {volume} {21}},\
		\bibinfo {pages} {391} (\bibinfo {year} {1973})}\BibitemShut {NoStop}%
	\bibitem [{\citenamefont {Larsson}\ and\ \citenamefont
		{Stenflo}(1973)}]{Larsson1973three}%
	\BibitemOpen
	\bibfield  {author} {\bibinfo {author} {\bibfnamefont {J.}~\bibnamefont
			{Larsson}}\ and\ \bibinfo {author} {\bibfnamefont {L.}~\bibnamefont
			{Stenflo}},\ }\bibfield  {title} {\bibinfo {title} {Three-wave interactions
			in magnetized plasmas},\ }\href@noop {} {\bibfield  {journal} {\bibinfo
			{journal} {Beitr{\"a}ge aus der Plasmaphysik}\ }\textbf {\bibinfo {volume}
			{13}},\ \bibinfo {pages} {169} (\bibinfo {year} {1973})}\BibitemShut
	{NoStop}%
	\bibitem [{\citenamefont {Stenflo}(1970{\natexlab{a}})}]{Stenflo70}%
	\BibitemOpen
	\bibfield  {author} {\bibinfo {author} {\bibfnamefont {L.}~\bibnamefont
			{Stenflo}},\ }\bibfield  {title} {\bibinfo {title} {Kinetic theory of
			three-wave interaction in a magnetied plasma},\ }\href@noop {} {\bibfield
		{journal} {\bibinfo  {journal} {Journal of Plasma Physics}\ }\textbf
		{\bibinfo {volume} {4}},\ \bibinfo {pages} {585} (\bibinfo {year}
		{1970}{\natexlab{a}})}\BibitemShut {NoStop}%
	\bibitem [{\citenamefont {Stenflo}(1970{\natexlab{b}})}]{stenflo1970effect}%
	\BibitemOpen
	\bibfield  {author} {\bibinfo {author} {\bibfnamefont {L.}~\bibnamefont
			{Stenflo}},\ }\bibfield  {title} {\bibinfo {title} {Effect of collisions on
			resonant three-wave interaction},\ }\href@noop {} {\bibfield  {journal}
		{\bibinfo  {journal} {Plasma Physics}\ }\textbf {\bibinfo {volume} {12}},\
		\bibinfo {pages} {509} (\bibinfo {year} {1970}{\natexlab{b}})}\BibitemShut
	{NoStop}%
	\bibitem [{\citenamefont {Shukla}\ and\ \citenamefont
		{Stenflo}(2002)}]{shukla2002modulational}%
	\BibitemOpen
	\bibfield  {author} {\bibinfo {author} {\bibfnamefont {P.~K.}\ \bibnamefont
			{Shukla}}\ and\ \bibinfo {author} {\bibfnamefont {L.}~\bibnamefont
			{Stenflo}},\ }\bibfield  {title} {\bibinfo {title} {Modulational
			instabilities of electromagnetic waves in a collision-dominated dust-electron
			plasma},\ }\href@noop {} {\bibfield  {journal} {\bibinfo  {journal} {IEEE
				transactions on plasma science}\ }\textbf {\bibinfo {volume} {29}},\ \bibinfo
		{pages} {267} (\bibinfo {year} {2002})}\BibitemShut {NoStop}%
	\bibitem [{\citenamefont {Larsson}\ \emph {et~al.}(1976)\citenamefont
		{Larsson}, \citenamefont {Stenflo},\ and\ \citenamefont
		{Tegeback}}]{Larsson76}%
	\BibitemOpen
	\bibfield  {author} {\bibinfo {author} {\bibfnamefont {J.}~\bibnamefont
			{Larsson}}, \bibinfo {author} {\bibfnamefont {L.}~\bibnamefont {Stenflo}},\
		and\ \bibinfo {author} {\bibfnamefont {R.}~\bibnamefont {Tegeback}},\
	}\bibfield  {title} {\bibinfo {title} {Enhanced fluctuations in a magnetized
			plasma due to the presence of an electromagnetic wave},\ }\href@noop {}
	{\bibfield  {journal} {\bibinfo  {journal} {Journal of Plasma Physics}\
		}\textbf {\bibinfo {volume} {16}},\ \bibinfo {pages} {37} (\bibinfo {year}
		{1976})}\BibitemShut {NoStop}%
	\bibitem [{\citenamefont {Stenflo}(2004)}]{Stenflo_2004}%
	\BibitemOpen
	\bibfield  {author} {\bibinfo {author} {\bibfnamefont {L.}~\bibnamefont
			{Stenflo}},\ }\bibfield  {title} {\bibinfo {title} {Comments on stimulated
			electromagnetic emissions in the ionospheric plasma},\ }\href
	{https://doi.org/10.1238/Physica.Topical.107a00262} {\bibfield  {journal}
		{\bibinfo  {journal} {Physica Scripta}\ }\textbf {\bibinfo {volume} {2004}},\
		\bibinfo {pages} {262} (\bibinfo {year} {2004})}\BibitemShut {NoStop}%
	\bibitem [{\citenamefont {Kaufman}\ and\ \citenamefont
		{Stenflo}(1979)}]{Kaufman_1979}%
	\BibitemOpen
	\bibfield  {author} {\bibinfo {author} {\bibfnamefont {A.~N.}\ \bibnamefont
			{Kaufman}}\ and\ \bibinfo {author} {\bibfnamefont {L.}~\bibnamefont
			{Stenflo}},\ }\bibfield  {title} {\bibinfo {title} {Wave coupling in cold
			nonuniform magnetoplasma},\ }\href
	{https://doi.org/10.1088/0031-8949/19/5-6/005} {\bibfield  {journal}
		{\bibinfo  {journal} {Physica Scripta}\ }\textbf {\bibinfo {volume} {19}},\
		\bibinfo {pages} {523} (\bibinfo {year} {1979})}\BibitemShut {NoStop}%
	\bibitem [{\citenamefont {Linden}\ \emph {et~al.}(1982)\citenamefont {Linden},
		\citenamefont {Larsson},\ and\ \citenamefont {Stenflo}}]{linden1982three}%
	\BibitemOpen
	\bibfield  {author} {\bibinfo {author} {\bibfnamefont {T.}~\bibnamefont
			{Linden}}, \bibinfo {author} {\bibfnamefont {J.}~\bibnamefont {Larsson}},\
		and\ \bibinfo {author} {\bibfnamefont {L.}~\bibnamefont {Stenflo}},\
	}\bibfield  {title} {\bibinfo {title} {Three-wave interaction in plasmas with
			sharp boundaries},\ }\href@noop {} {\bibfield  {journal} {\bibinfo  {journal}
			{Plasma Physics}\ }\textbf {\bibinfo {volume} {24}},\ \bibinfo {pages} {1177}
		(\bibinfo {year} {1982})}\BibitemShut {NoStop}%
	\bibitem [{\citenamefont {Lindgren}\ \emph {et~al.}(1985)\citenamefont
		{Lindgren}, \citenamefont {Stenflo}, \citenamefont {Kostov},\ and\
		\citenamefont {Zhelyazkov}}]{Lindgren_Stenflo_Kostov_Zhelyazkov_1985}%
	\BibitemOpen
	\bibfield  {author} {\bibinfo {author} {\bibfnamefont {T.}~\bibnamefont
			{Lindgren}}, \bibinfo {author} {\bibfnamefont {L.}~\bibnamefont {Stenflo}},
		\bibinfo {author} {\bibfnamefont {N.}~\bibnamefont {Kostov}},\ and\ \bibinfo
		{author} {\bibfnamefont {I.}~\bibnamefont {Zhelyazkov}},\ }\bibfield  {title}
	{\bibinfo {title} {Three-wave interaction in a cold plasma column},\ }\href
	{https://doi.org/10.1017/S0022377800002981} {\bibfield  {journal} {\bibinfo
			{journal} {Journal of Plasma Physics}\ }\textbf {\bibinfo {volume} {34}},\
		\bibinfo {pages} {427–434} (\bibinfo {year} {1985})}\BibitemShut {NoStop}%
	\bibitem [{\citenamefont {Axelsson}\ \emph {et~al.}(1996)\citenamefont
		{Axelsson}, \citenamefont {Larsson},\ and\ \citenamefont
		{Stenflo}}]{axelsson1996nonlinear}%
	\BibitemOpen
	\bibfield  {author} {\bibinfo {author} {\bibfnamefont {P.}~\bibnamefont
			{Axelsson}}, \bibinfo {author} {\bibfnamefont {J.}~\bibnamefont {Larsson}},\
		and\ \bibinfo {author} {\bibfnamefont {L.}~\bibnamefont {Stenflo}},\
	}\bibfield  {title} {\bibinfo {title} {Nonlinear interaction between acoustic
			gravity waves},\ }in\ \href@noop {} {\emph {\bibinfo {booktitle} {Annales
				Geophysicae}}},\ Vol.~\bibinfo {volume} {14}\ (\bibinfo {organization}
	{Springer Verlag G{\"o}ttingen, Germany},\ \bibinfo {year} {1996})\ pp.\
	\bibinfo {pages} {304--308}\BibitemShut {NoStop}%
	\bibitem [{\citenamefont {Stenflo}\ and\ \citenamefont
		{Shukla}(2009{\natexlab{a}})}]{STENFLO_SHUKLA_2009}%
	\BibitemOpen
	\bibfield  {author} {\bibinfo {author} {\bibfnamefont {L.}~\bibnamefont
			{Stenflo}}\ and\ \bibinfo {author} {\bibfnamefont {P.~K.}\ \bibnamefont
			{Shukla}},\ }\bibfield  {title} {\bibinfo {title} {Nonlinear
			acoustic–gravity waves},\ }\href
	{https://doi.org/10.1017/S0022377809007892} {\bibfield  {journal} {\bibinfo
			{journal} {Journal of Plasma Physics}\ }\textbf {\bibinfo {volume} {75}},\
		\bibinfo {pages} {841–847} (\bibinfo {year}
		{2009}{\natexlab{a}})}\BibitemShut {NoStop}%
	\bibitem [{\citenamefont {Vladimirov}\ and\ \citenamefont
		{Stenflo}(1997)}]{vladimirov1997three}%
	\BibitemOpen
	\bibfield  {author} {\bibinfo {author} {\bibfnamefont {S.~V.}\ \bibnamefont
			{Vladimirov}}\ and\ \bibinfo {author} {\bibfnamefont {L.}~\bibnamefont
			{Stenflo}},\ }\bibfield  {title} {\bibinfo {title} {Three-wave processes in a
			turbulent nonstationary plasma},\ }\href@noop {} {\bibfield  {journal}
		{\bibinfo  {journal} {Physics of Plasmas}\ }\textbf {\bibinfo {volume} {4}},\
		\bibinfo {pages} {1249} (\bibinfo {year} {1997})}\BibitemShut {NoStop}%
	\bibitem [{\citenamefont {Brodin}\ and\ \citenamefont
		{Stenflo}(1989)}]{brodin1989three}%
	\BibitemOpen
	\bibfield  {author} {\bibinfo {author} {\bibfnamefont {G.}~\bibnamefont
			{Brodin}}\ and\ \bibinfo {author} {\bibfnamefont {L.}~\bibnamefont
			{Stenflo}},\ }\bibfield  {title} {\bibinfo {title} {Three-wave interaction
			between transverse and longitudinal waves},\ }\href@noop {} {\bibfield
		{journal} {\bibinfo  {journal} {Journal of plasma physics}\ }\textbf
		{\bibinfo {volume} {42}},\ \bibinfo {pages} {187} (\bibinfo {year}
		{1989})}\BibitemShut {NoStop}%
	\bibitem [{\citenamefont {Brodin}\ \emph {et~al.}(2007)\citenamefont {Brodin},
		\citenamefont {Stenflo},\ and\ \citenamefont {Shukla}}]{brodin2007nonlinear}%
	\BibitemOpen
	\bibfield  {author} {\bibinfo {author} {\bibfnamefont {G.}~\bibnamefont
			{Brodin}}, \bibinfo {author} {\bibfnamefont {L.}~\bibnamefont {Stenflo}},\
		and\ \bibinfo {author} {\bibfnamefont {P.~K.}\ \bibnamefont {Shukla}},\
	}\bibfield  {title} {\bibinfo {title} {Nonlinear interactions between three
			inertial alfv{\'e}n waves},\ }\href@noop {} {\bibfield  {journal} {\bibinfo
			{journal} {Journal of plasma physics}\ }\textbf {\bibinfo {volume} {73}},\
		\bibinfo {pages} {9} (\bibinfo {year} {2007})}\BibitemShut {NoStop}%
	\bibitem [{\citenamefont {Brodin}\ and\ \citenamefont
		{Stenflo}(2016)}]{Brodin2016new}%
	\BibitemOpen
	\bibfield  {author} {\bibinfo {author} {\bibfnamefont {G.}~\bibnamefont
			{Brodin}}\ and\ \bibinfo {author} {\bibfnamefont {L.}~\bibnamefont
			{Stenflo}},\ }\bibfield  {title} {\bibinfo {title} {A new decay channel for
			upper-hybrid waves},\ }\href@noop {} {\bibfield  {journal} {\bibinfo
			{journal} {Physica Scripta}\ }\textbf {\bibinfo {volume} {91}},\ \bibinfo
		{pages} {104005} (\bibinfo {year} {2016})}\BibitemShut {NoStop}%
	\bibitem [{\citenamefont {Larsson}\ and\ \citenamefont
		{Stenflo}(1976)}]{Larsson1976parametric}%
	\BibitemOpen
	\bibfield  {author} {\bibinfo {author} {\bibfnamefont {J.}~\bibnamefont
			{Larsson}}\ and\ \bibinfo {author} {\bibfnamefont {L.}~\bibnamefont
			{Stenflo}},\ }\bibfield  {title} {\bibinfo {title} {Parametric instabilities
			of waves in magnetized plasmas},\ }\href@noop {} {\bibfield  {journal}
		{\bibinfo  {journal} {Beitraege Plasmaphysik}\ }\textbf {\bibinfo {volume}
			{16}},\ \bibinfo {pages} {79} (\bibinfo {year} {1976})}\BibitemShut {NoStop}%
	\bibitem [{\citenamefont {Stenflo}(1994)}]{Stenflo94}%
	\BibitemOpen
	\bibfield  {author} {\bibinfo {author} {\bibfnamefont {L.}~\bibnamefont
			{Stenflo}},\ }\bibfield  {title} {\bibinfo {title} {Resonant three-wave
			interactions in plasmas},\ }\href@noop {} {\bibfield  {journal} {\bibinfo
			{journal} {Physica Scripta}\ }\textbf {\bibinfo {volume} {1994}},\ \bibinfo
		{pages} {15} (\bibinfo {year} {1994})}\BibitemShut {NoStop}%
	\bibitem [{\citenamefont {Stenflo}\ and\ \citenamefont
		{Brodin}(2006)}]{Stenflo2006three}%
	\BibitemOpen
	\bibfield  {author} {\bibinfo {author} {\bibfnamefont {L.}~\bibnamefont
			{Stenflo}}\ and\ \bibinfo {author} {\bibfnamefont {G.}~\bibnamefont
			{Brodin}},\ }\bibfield  {title} {\bibinfo {title} {The three-wave coupling
			coefficients for a cold magnetized plasma},\ }\href@noop {} {\bibfield
		{journal} {\bibinfo  {journal} {Journal of plasma physics}\ }\textbf
		{\bibinfo {volume} {72}},\ \bibinfo {pages} {143} (\bibinfo {year}
		{2006})}\BibitemShut {NoStop}%
	\bibitem [{\citenamefont {Brodin}\ and\ \citenamefont
		{Stenflo}(2012)}]{Brodin12}%
	\BibitemOpen
	\bibfield  {author} {\bibinfo {author} {\bibfnamefont {G.}~\bibnamefont
			{Brodin}}\ and\ \bibinfo {author} {\bibfnamefont {L.}~\bibnamefont
			{Stenflo}},\ }\bibfield  {title} {\bibinfo {title} {Three-wave coupling
			coefficients for a magnetized plasma},\ }\href@noop {} {\bibfield  {journal}
		{\bibinfo  {journal} {Physica Scripta}\ }\textbf {\bibinfo {volume} {85}},\
		\bibinfo {pages} {035504} (\bibinfo {year} {2012})}\BibitemShut {NoStop}%
	\bibitem [{\citenamefont {Stenflo}(1990)}]{Stenflo_1990}%
	\BibitemOpen
	\bibfield  {author} {\bibinfo {author} {\bibfnamefont {L.}~\bibnamefont
			{Stenflo}},\ }\bibfield  {title} {\bibinfo {title} {Stimulated scattering of
			large amplitude waves in the ionosphere},\ }\href
	{https://doi.org/10.1088/0031-8949/1990/T30/022} {\bibfield  {journal}
		{\bibinfo  {journal} {Physica Scripta}\ }\textbf {\bibinfo {volume} {1990}},\
		\bibinfo {pages} {166} (\bibinfo {year} {1990})}\BibitemShut {NoStop}%
	\bibitem [{\citenamefont {Stenflo}(1981{\natexlab{b}})}]{Stenflo1981self}%
	\BibitemOpen
	\bibfield  {author} {\bibinfo {author} {\bibfnamefont {L.}~\bibnamefont
			{Stenflo}},\ }\bibfield  {title} {\bibinfo {title} {Self-consistent vlasov
			description of a magnetized plasma in a large amplitude circularly polarized
			wave},\ }\href@noop {} {\bibfield  {journal} {\bibinfo  {journal} {Physica
				Scripta}\ }\textbf {\bibinfo {volume} {23}},\ \bibinfo {pages} {779}
		(\bibinfo {year} {1981}{\natexlab{b}})}\BibitemShut {NoStop}%
	\bibitem [{\citenamefont {Brodin}\ and\ \citenamefont
		{Stenflo}(2014)}]{brodin2014wave}%
	\BibitemOpen
	\bibfield  {author} {\bibinfo {author} {\bibfnamefont {G.}~\bibnamefont
			{Brodin}}\ and\ \bibinfo {author} {\bibfnamefont {L.}~\bibnamefont
			{Stenflo}},\ }\bibfield  {title} {\bibinfo {title} {Wave generation in a warm
			magnetized multi-component plasma},\ }\href@noop {} {\bibfield  {journal}
		{\bibinfo  {journal} {Contributions to Plasma Physics}\ }\textbf {\bibinfo
			{volume} {54}},\ \bibinfo {pages} {623} (\bibinfo {year} {2014})}\BibitemShut
	{NoStop}%
	\bibitem [{\citenamefont {Stenflo}\ and\ \citenamefont
		{Shukla}(2009{\natexlab{b}})}]{Stenflo2009wave}%
	\BibitemOpen
	\bibfield  {author} {\bibinfo {author} {\bibfnamefont {L.}~\bibnamefont
			{Stenflo}}\ and\ \bibinfo {author} {\bibfnamefont {P.}~\bibnamefont
			{Shukla}},\ }\bibfield  {title} {\bibinfo {title} {Wave-wave interactions in
			plasmas},\ }in\ \href@noop {} {\emph {\bibinfo {booktitle} {AIP Conference
				Proceedings}}},\ Vol.\ \bibinfo {volume} {1177}\ (\bibinfo {organization}
	{American Institute of Physics},\ \bibinfo {year} {2009})\ pp.\ \bibinfo
	{pages} {4--9}\BibitemShut {NoStop}%
	\bibitem [{\citenamefont {Moody}\ and\ \citenamefont {Shi}(2025)}]{Moody25}%
	\BibitemOpen
	\bibfield  {author} {\bibinfo {author} {\bibfnamefont {J.~D.}\ \bibnamefont
			{Moody}}\ and\ \bibinfo {author} {\bibfnamefont {Y.}~\bibnamefont {Shi}},\
	}\bibfield  {title} {\bibinfo {title} {Vlasov theory of magnetized cross-beam
			energy transfer in a high energy densityplasma},\ }\href@noop {} {\bibfield
		{journal} {\bibinfo  {journal} {To be submitted}\ } (\bibinfo {year}
		{2025})}\BibitemShut {NoStop}%
	\bibitem [{\citenamefont {Randall}\ \emph {et~al.}(1981)\citenamefont
		{Randall}, \citenamefont {Albritton},\ and\ \citenamefont
		{Thomson}}]{Randall81}%
	\BibitemOpen
	\bibfield  {author} {\bibinfo {author} {\bibfnamefont {C.~J.}\ \bibnamefont
			{Randall}}, \bibinfo {author} {\bibfnamefont {J.~R.}\ \bibnamefont
			{Albritton}},\ and\ \bibinfo {author} {\bibfnamefont {J.~J.}\ \bibnamefont
			{Thomson}},\ }\bibfield  {title} {\bibinfo {title} {Theory and simulation of
			stimulated {Brillouin} scatter excited by nonabsorbed light in laser fusion
			systems},\ }\href {https://doi.org/10.1063/1.863551} {\bibfield  {journal}
		{\bibinfo  {journal} {The Physics of Fluids}\ }\textbf {\bibinfo {volume}
			{24}},\ \bibinfo {pages} {1474} (\bibinfo {year} {1981})}\BibitemShut
	{NoStop}%
	\bibitem [{\citenamefont {Michel}(2023)}]{Michel23}%
	\BibitemOpen
	\bibfield  {author} {\bibinfo {author} {\bibfnamefont {P.}~\bibnamefont
			{Michel}},\ }\href@noop {} {\emph {\bibinfo {title} {Introduction to
				laser-plasma interactions}}}\ (\bibinfo  {publisher} {Springer Nature},\
	\bibinfo {year} {2023})\BibitemShut {NoStop}%
	\bibitem [{\citenamefont {Seka}\ \emph {et~al.}(2008)\citenamefont {Seka},
		\citenamefont {Edgell}, \citenamefont {Knauer}, \citenamefont {Myatt},
		\citenamefont {Maximov}, \citenamefont {Short}, \citenamefont {Sangster},
		\citenamefont {Stoeckl}, \citenamefont {Bahr}, \citenamefont {Craxton},
		\citenamefont {Delettrez}, \citenamefont {Goncharov}, \citenamefont
		{Igumenshchev},\ and\ \citenamefont {Shvarts}}]{Seka08}%
	\BibitemOpen
	\bibfield  {author} {\bibinfo {author} {\bibfnamefont {W.}~\bibnamefont
			{Seka}}, \bibinfo {author} {\bibfnamefont {D.~H.}\ \bibnamefont {Edgell}},
		\bibinfo {author} {\bibfnamefont {J.~P.}\ \bibnamefont {Knauer}}, \bibinfo
		{author} {\bibfnamefont {J.~F.}\ \bibnamefont {Myatt}}, \bibinfo {author}
		{\bibfnamefont {A.~V.}\ \bibnamefont {Maximov}}, \bibinfo {author}
		{\bibfnamefont {R.~W.}\ \bibnamefont {Short}}, \bibinfo {author}
		{\bibfnamefont {T.~C.}\ \bibnamefont {Sangster}}, \bibinfo {author}
		{\bibfnamefont {C.}~\bibnamefont {Stoeckl}}, \bibinfo {author} {\bibfnamefont
			{R.~E.}\ \bibnamefont {Bahr}}, \bibinfo {author} {\bibfnamefont {R.~S.}\
			\bibnamefont {Craxton}}, \bibinfo {author} {\bibfnamefont {J.~A.}\
			\bibnamefont {Delettrez}}, \bibinfo {author} {\bibfnamefont {V.~N.}\
			\bibnamefont {Goncharov}}, \bibinfo {author} {\bibfnamefont {I.~V.}\
			\bibnamefont {Igumenshchev}},\ and\ \bibinfo {author} {\bibfnamefont
			{D.}~\bibnamefont {Shvarts}},\ }\bibfield  {title} {\bibinfo {title}
		{Time-resolved absorption in cryogenic and room-temperature direct-drive
			implosion},\ }\href {https://doi.org/10.1063/1.2898405} {\bibfield  {journal}
		{\bibinfo  {journal} {Physics of Plasmas}\ }\textbf {\bibinfo {volume}
			{15}},\ \bibinfo {pages} {056312} (\bibinfo {year} {2008})}\BibitemShut
	{NoStop}%
	\bibitem [{\citenamefont {Froula}\ \emph {et~al.}(2012)\citenamefont {Froula},
		\citenamefont {Igumenshchev}, \citenamefont {Michel}, \citenamefont {Edgell},
		\citenamefont {Follett}, \citenamefont {Glebov}, \citenamefont {Goncharov},
		\citenamefont {Kwiatkowski}, \citenamefont {Marshall}, \citenamefont {Radha},
		\citenamefont {Seka}, \citenamefont {Sorce}, \citenamefont {Stagnitto},
		\citenamefont {Stoeckl},\ and\ \citenamefont {Sangster}}]{Froula12}%
	\BibitemOpen
	\bibfield  {author} {\bibinfo {author} {\bibfnamefont {D.~H.}\ \bibnamefont
			{Froula}}, \bibinfo {author} {\bibfnamefont {I.~V.}\ \bibnamefont
			{Igumenshchev}}, \bibinfo {author} {\bibfnamefont {D.~T.}\ \bibnamefont
			{Michel}}, \bibinfo {author} {\bibfnamefont {D.~H.}\ \bibnamefont {Edgell}},
		\bibinfo {author} {\bibfnamefont {R.}~\bibnamefont {Follett}}, \bibinfo
		{author} {\bibfnamefont {V.~Y.}\ \bibnamefont {Glebov}}, \bibinfo {author}
		{\bibfnamefont {V.~N.}\ \bibnamefont {Goncharov}}, \bibinfo {author}
		{\bibfnamefont {J.}~\bibnamefont {Kwiatkowski}}, \bibinfo {author}
		{\bibfnamefont {F.~J.}\ \bibnamefont {Marshall}}, \bibinfo {author}
		{\bibfnamefont {P.~B.}\ \bibnamefont {Radha}}, \bibinfo {author}
		{\bibfnamefont {W.}~\bibnamefont {Seka}}, \bibinfo {author} {\bibfnamefont
			{C.}~\bibnamefont {Sorce}}, \bibinfo {author} {\bibfnamefont
			{S.}~\bibnamefont {Stagnitto}}, \bibinfo {author} {\bibfnamefont
			{C.}~\bibnamefont {Stoeckl}},\ and\ \bibinfo {author} {\bibfnamefont {T.~C.}\
			\bibnamefont {Sangster}},\ }\bibfield  {title} {\bibinfo {title} {Increasing
			hydrodynamic efficiency by reducing cross-beam energy transfer in
			direct-drive-implosion experiments},\ }\href
	{https://doi.org/10.1103/PhysRevLett.108.125003} {\bibfield  {journal}
		{\bibinfo  {journal} {Physical Review Letters}\ }\textbf {\bibinfo {volume}
			{108}},\ \bibinfo {pages} {125003} (\bibinfo {year} {2012})}\BibitemShut
	{NoStop}%
	\bibitem [{\citenamefont {Marozas}\ \emph {et~al.}(2018)\citenamefont
		{Marozas}, \citenamefont {Hohenberger}, \citenamefont {Rosenberg},
		\citenamefont {Turnbull}, \citenamefont {Collins}, \citenamefont {Radha},
		\citenamefont {McKenty}, \citenamefont {Zuegel}, \citenamefont {Marshall},
		\citenamefont {Regan}, \citenamefont {Sangster}, \citenamefont {Seka},
		\citenamefont {Campbell}, \citenamefont {Goncharov}, \citenamefont {Bowers},
		\citenamefont {Di~Nicola}, \citenamefont {Erbert}, \citenamefont {MacGowan},
		\citenamefont {Pelz},\ and\ \citenamefont {Yang}}]{Marozas18}%
	\BibitemOpen
	\bibfield  {author} {\bibinfo {author} {\bibfnamefont {J.~A.}\ \bibnamefont
			{Marozas}}, \bibinfo {author} {\bibfnamefont {M.}~\bibnamefont
			{Hohenberger}}, \bibinfo {author} {\bibfnamefont {M.~J.}\ \bibnamefont
			{Rosenberg}}, \bibinfo {author} {\bibfnamefont {D.}~\bibnamefont {Turnbull}},
		\bibinfo {author} {\bibfnamefont {T.~J.~B.}\ \bibnamefont {Collins}},
		\bibinfo {author} {\bibfnamefont {P.~B.}\ \bibnamefont {Radha}}, \bibinfo
		{author} {\bibfnamefont {P.~W.}\ \bibnamefont {McKenty}}, \bibinfo {author}
		{\bibfnamefont {J.~D.}\ \bibnamefont {Zuegel}}, \bibinfo {author}
		{\bibfnamefont {F.~J.}\ \bibnamefont {Marshall}}, \bibinfo {author}
		{\bibfnamefont {S.~P.}\ \bibnamefont {Regan}}, \bibinfo {author}
		{\bibfnamefont {T.~C.}\ \bibnamefont {Sangster}}, \bibinfo {author}
		{\bibfnamefont {W.}~\bibnamefont {Seka}}, \bibinfo {author} {\bibfnamefont
			{E.~M.}\ \bibnamefont {Campbell}}, \bibinfo {author} {\bibfnamefont {V.~N.}\
			\bibnamefont {Goncharov}}, \bibinfo {author} {\bibfnamefont {M.~W.}\
			\bibnamefont {Bowers}}, \bibinfo {author} {\bibfnamefont {J.-M.~G.}\
			\bibnamefont {Di~Nicola}}, \bibinfo {author} {\bibfnamefont {G.}~\bibnamefont
			{Erbert}}, \bibinfo {author} {\bibfnamefont {B.~J.}\ \bibnamefont
			{MacGowan}}, \bibinfo {author} {\bibfnamefont {L.~J.}\ \bibnamefont {Pelz}},\
		and\ \bibinfo {author} {\bibfnamefont {S.~T.}\ \bibnamefont {Yang}},\
	}\bibfield  {title} {\bibinfo {title} {First observation of cross-beam energy
			transfer mitigation for direct-drive inertial confinement fusion implosions
			using wavelength detuning at the {National Ignition Facility}},\ }\href
	{https://doi.org/10.1103/PhysRevLett.120.085001} {\bibfield  {journal}
		{\bibinfo  {journal} {Physical Review Letters}\ }\textbf {\bibinfo {volume}
			{120}},\ \bibinfo {pages} {085001} (\bibinfo {year} {2018})}\BibitemShut
	{NoStop}%
	\bibitem [{\citenamefont {Michel}\ \emph {et~al.}(2011)\citenamefont {Michel},
		\citenamefont {Divol}, \citenamefont {Town}, \citenamefont {Rosen},
		\citenamefont {Callahan}, \citenamefont {Meezan}, \citenamefont {Schneider},
		\citenamefont {Kyrala}, \citenamefont {Moody}, \citenamefont {Dewald},
		\citenamefont {Widmann}, \citenamefont {Bond}, \citenamefont {Kline},
		\citenamefont {Thomas}, \citenamefont {Dixit}, \citenamefont {Williams},
		\citenamefont {Hinkel}, \citenamefont {Berger}, \citenamefont {Landen},
		\citenamefont {Edwards}, \citenamefont {MacGowan}, \citenamefont {Lindl},
		\citenamefont {Haynam}, \citenamefont {Suter}, \citenamefont {Glenzer},\ and\
		\citenamefont {Moses}}]{Michel11}%
	\BibitemOpen
	\bibfield  {author} {\bibinfo {author} {\bibfnamefont {P.}~\bibnamefont
			{Michel}}, \bibinfo {author} {\bibfnamefont {L.}~\bibnamefont {Divol}},
		\bibinfo {author} {\bibfnamefont {R.~P.~J.}\ \bibnamefont {Town}}, \bibinfo
		{author} {\bibfnamefont {M.~D.}\ \bibnamefont {Rosen}}, \bibinfo {author}
		{\bibfnamefont {D.~A.}\ \bibnamefont {Callahan}}, \bibinfo {author}
		{\bibfnamefont {N.~B.}\ \bibnamefont {Meezan}}, \bibinfo {author}
		{\bibfnamefont {M.~B.}\ \bibnamefont {Schneider}}, \bibinfo {author}
		{\bibfnamefont {G.~A.}\ \bibnamefont {Kyrala}}, \bibinfo {author}
		{\bibfnamefont {J.~D.}\ \bibnamefont {Moody}}, \bibinfo {author}
		{\bibfnamefont {E.~L.}\ \bibnamefont {Dewald}}, \bibinfo {author}
		{\bibfnamefont {K.}~\bibnamefont {Widmann}}, \bibinfo {author} {\bibfnamefont
			{E.}~\bibnamefont {Bond}}, \bibinfo {author} {\bibfnamefont {J.~L.}\
			\bibnamefont {Kline}}, \bibinfo {author} {\bibfnamefont {C.~A.}\ \bibnamefont
			{Thomas}}, \bibinfo {author} {\bibfnamefont {S.}~\bibnamefont {Dixit}},
		\bibinfo {author} {\bibfnamefont {E.~A.}\ \bibnamefont {Williams}}, \bibinfo
		{author} {\bibfnamefont {D.~E.}\ \bibnamefont {Hinkel}}, \bibinfo {author}
		{\bibfnamefont {R.~L.}\ \bibnamefont {Berger}}, \bibinfo {author}
		{\bibfnamefont {O.~L.}\ \bibnamefont {Landen}}, \bibinfo {author}
		{\bibfnamefont {M.~J.}\ \bibnamefont {Edwards}}, \bibinfo {author}
		{\bibfnamefont {B.~J.}\ \bibnamefont {MacGowan}}, \bibinfo {author}
		{\bibfnamefont {J.~D.}\ \bibnamefont {Lindl}}, \bibinfo {author}
		{\bibfnamefont {C.}~\bibnamefont {Haynam}}, \bibinfo {author} {\bibfnamefont
			{L.~J.}\ \bibnamefont {Suter}}, \bibinfo {author} {\bibfnamefont {S.~H.}\
			\bibnamefont {Glenzer}},\ and\ \bibinfo {author} {\bibfnamefont
			{E.}~\bibnamefont {Moses}},\ }\bibfield  {title} {\bibinfo {title}
		{Three-wavelength scheme to optimize hohlraum coupling on the {National
				Ignition Facility}},\ }\href {https://doi.org/10.1103/PhysRevE.83.046409}
	{\bibfield  {journal} {\bibinfo  {journal} {Physical Review E}\ }\textbf
		{\bibinfo {volume} {83}},\ \bibinfo {pages} {046409} (\bibinfo {year}
		{2011})}\BibitemShut {NoStop}%
	\bibitem [{\citenamefont {Moody}\ \emph {et~al.}(2012)\citenamefont {Moody},
		\citenamefont {Michel}, \citenamefont {Divol}, \citenamefont {Berger},
		\citenamefont {Bond}, \citenamefont {Bradley}, \citenamefont {Callahan},
		\citenamefont {Dewald}, \citenamefont {Dixit}, \citenamefont {Edwards} \emph
		{et~al.}}]{Moody12}%
	\BibitemOpen
	\bibfield  {author} {\bibinfo {author} {\bibfnamefont {J.}~\bibnamefont
			{Moody}}, \bibinfo {author} {\bibfnamefont {P.}~\bibnamefont {Michel}},
		\bibinfo {author} {\bibfnamefont {L.}~\bibnamefont {Divol}}, \bibinfo
		{author} {\bibfnamefont {R.}~\bibnamefont {Berger}}, \bibinfo {author}
		{\bibfnamefont {E.}~\bibnamefont {Bond}}, \bibinfo {author} {\bibfnamefont
			{D.}~\bibnamefont {Bradley}}, \bibinfo {author} {\bibfnamefont
			{D.}~\bibnamefont {Callahan}}, \bibinfo {author} {\bibfnamefont
			{E.}~\bibnamefont {Dewald}}, \bibinfo {author} {\bibfnamefont
			{S.}~\bibnamefont {Dixit}}, \bibinfo {author} {\bibfnamefont
			{M.}~\bibnamefont {Edwards}}, \emph {et~al.},\ }\bibfield  {title} {\bibinfo
		{title} {Multistep redirection by cross-beam power transfer of
			ultrahigh-power lasers in a plasma},\ }\href@noop {} {\bibfield  {journal}
		{\bibinfo  {journal} {Nature Physics}\ }\textbf {\bibinfo {volume} {8}},\
		\bibinfo {pages} {344} (\bibinfo {year} {2012})}\BibitemShut {NoStop}%
	\bibitem [{\citenamefont {Pickworth}\ \emph {et~al.}(2020)\citenamefont
		{Pickworth}, \citenamefont {Döppner}, \citenamefont {Hinkel}, \citenamefont
		{Ralph}, \citenamefont {Bachmann}, \citenamefont {Masse}, \citenamefont
		{Divol}, \citenamefont {Benedetti}, \citenamefont {Celliers}, \citenamefont
		{Chen}, \citenamefont {Hohenberger}, \citenamefont {Khan}, \citenamefont
		{Landen}, \citenamefont {Lemos}, \citenamefont {MacGowan}, \citenamefont
		{Mariscal}, \citenamefont {Michel}, \citenamefont {Millot}, \citenamefont
		{Moore}, \citenamefont {Park}, \citenamefont {Schneider}, \citenamefont
		{Callahan},\ and\ \citenamefont {Hurricane}}]{Pickworth20}%
	\BibitemOpen
	\bibfield  {author} {\bibinfo {author} {\bibfnamefont {L.~A.}\ \bibnamefont
			{Pickworth}}, \bibinfo {author} {\bibfnamefont {T.}~\bibnamefont {Döppner}},
		\bibinfo {author} {\bibfnamefont {D.~E.}\ \bibnamefont {Hinkel}}, \bibinfo
		{author} {\bibfnamefont {J.~E.}\ \bibnamefont {Ralph}}, \bibinfo {author}
		{\bibfnamefont {B.}~\bibnamefont {Bachmann}}, \bibinfo {author}
		{\bibfnamefont {L.~P.}\ \bibnamefont {Masse}}, \bibinfo {author}
		{\bibfnamefont {L.}~\bibnamefont {Divol}}, \bibinfo {author} {\bibfnamefont
			{L.~R.}\ \bibnamefont {Benedetti}}, \bibinfo {author} {\bibfnamefont {P.~M.}\
			\bibnamefont {Celliers}}, \bibinfo {author} {\bibfnamefont {H.}~\bibnamefont
			{Chen}}, \bibinfo {author} {\bibfnamefont {M.}~\bibnamefont {Hohenberger}},
		\bibinfo {author} {\bibfnamefont {S.~F.}\ \bibnamefont {Khan}}, \bibinfo
		{author} {\bibfnamefont {O.~L.}\ \bibnamefont {Landen}}, \bibinfo {author}
		{\bibfnamefont {N.}~\bibnamefont {Lemos}}, \bibinfo {author} {\bibfnamefont
			{B.~J.}\ \bibnamefont {MacGowan}}, \bibinfo {author} {\bibfnamefont {D.~A.}\
			\bibnamefont {Mariscal}}, \bibinfo {author} {\bibfnamefont {P.~A.}\
			\bibnamefont {Michel}}, \bibinfo {author} {\bibfnamefont {M.}~\bibnamefont
			{Millot}}, \bibinfo {author} {\bibfnamefont {A.~S.}\ \bibnamefont {Moore}},
		\bibinfo {author} {\bibfnamefont {J.}~\bibnamefont {Park}}, \bibinfo {author}
		{\bibfnamefont {M.~B.}\ \bibnamefont {Schneider}}, \bibinfo {author}
		{\bibfnamefont {D.~A.}\ \bibnamefont {Callahan}},\ and\ \bibinfo {author}
		{\bibfnamefont {O.~A.}\ \bibnamefont {Hurricane}},\ }\bibfield  {title}
	{\bibinfo {title} {Application of cross-beam energy transfer to control drive
			symmetry in {ICF} implosions in low gas fill hohlraums at the national
			ignition facility},\ }\href {https://doi.org/10.1063/5.0004866} {\bibfield
		{journal} {\bibinfo  {journal} {Physics of Plasmas}\ }\textbf {\bibinfo
			{volume} {27}},\ \bibinfo {pages} {102702} (\bibinfo {year}
		{2020})}\BibitemShut {NoStop}%
	\bibitem [{\citenamefont {Arber}\ \emph {et~al.}(2015)\citenamefont {Arber},
		\citenamefont {Bennett}, \citenamefont {Brady}, \citenamefont
		{Lawrence-Douglas}, \citenamefont {Ramsay}, \citenamefont {Sircombe},
		\citenamefont {Gillies}, \citenamefont {Evans}, \citenamefont {Schmitz},
		\citenamefont {Bell},\ and\ \citenamefont {Ridgers}}]{Arber:2015hc}%
	\BibitemOpen
	\bibfield  {author} {\bibinfo {author} {\bibfnamefont {T.~D.}\ \bibnamefont
			{Arber}}, \bibinfo {author} {\bibfnamefont {K.}~\bibnamefont {Bennett}},
		\bibinfo {author} {\bibfnamefont {C.~S.}\ \bibnamefont {Brady}}, \bibinfo
		{author} {\bibfnamefont {A.}~\bibnamefont {Lawrence-Douglas}}, \bibinfo
		{author} {\bibfnamefont {M.~G.}\ \bibnamefont {Ramsay}}, \bibinfo {author}
		{\bibfnamefont {N.~J.}\ \bibnamefont {Sircombe}}, \bibinfo {author}
		{\bibfnamefont {P.}~\bibnamefont {Gillies}}, \bibinfo {author} {\bibfnamefont
			{R.~G.}\ \bibnamefont {Evans}}, \bibinfo {author} {\bibfnamefont
			{H.}~\bibnamefont {Schmitz}}, \bibinfo {author} {\bibfnamefont {A.~R.}\
			\bibnamefont {Bell}},\ and\ \bibinfo {author} {\bibfnamefont {C.~P.}\
			\bibnamefont {Ridgers}},\ }\bibfield  {title} {\bibinfo {title}
		{{Contemporary particle-in-cell approach to laser-plasma modelling}},\
	}\href@noop {} {\bibfield  {journal} {\bibinfo  {journal} {Plasma Physics and
				Controlled Fusion}\ }\textbf {\bibinfo {volume} {57}},\ \bibinfo {pages} {1}
		(\bibinfo {year} {2015})}\BibitemShut {NoStop}%
	\bibitem [{\citenamefont {Turnbull}\ \emph {et~al.}(2020)\citenamefont
		{Turnbull}, \citenamefont {Cola{\"\i}tis}, \citenamefont {Hansen},
		\citenamefont {Milder}, \citenamefont {Palastro}, \citenamefont {Katz},
		\citenamefont {Dorrer}, \citenamefont {Kruschwitz}, \citenamefont {Strozzi},\
		and\ \citenamefont {Froula}}]{Turnbull20}%
	\BibitemOpen
	\bibfield  {author} {\bibinfo {author} {\bibfnamefont {D.}~\bibnamefont
			{Turnbull}}, \bibinfo {author} {\bibfnamefont {A.}~\bibnamefont
			{Cola{\"\i}tis}}, \bibinfo {author} {\bibfnamefont {A.~M.}\ \bibnamefont
			{Hansen}}, \bibinfo {author} {\bibfnamefont {A.~L.}\ \bibnamefont {Milder}},
		\bibinfo {author} {\bibfnamefont {J.~P.}\ \bibnamefont {Palastro}}, \bibinfo
		{author} {\bibfnamefont {J.}~\bibnamefont {Katz}}, \bibinfo {author}
		{\bibfnamefont {C.}~\bibnamefont {Dorrer}}, \bibinfo {author} {\bibfnamefont
			{B.~E.}\ \bibnamefont {Kruschwitz}}, \bibinfo {author} {\bibfnamefont
			{D.~J.}\ \bibnamefont {Strozzi}},\ and\ \bibinfo {author} {\bibfnamefont
			{D.~H.}\ \bibnamefont {Froula}},\ }\bibfield  {title} {\bibinfo {title}
		{Impact of the langdon effect on crossed-beam energy transfer},\ }\href@noop
	{} {\bibfield  {journal} {\bibinfo  {journal} {Nature Physics}\ }\textbf
		{\bibinfo {volume} {16}},\ \bibinfo {pages} {181} (\bibinfo {year}
		{2020})}\BibitemShut {NoStop}%
	\bibitem [{\citenamefont {Hansen}\ \emph {et~al.}(2022)\citenamefont {Hansen},
		\citenamefont {Nguyen}, \citenamefont {Turnbull}, \citenamefont {Albright},
		\citenamefont {Follett}, \citenamefont {Huff}, \citenamefont {Katz},
		\citenamefont {Mastrosimone}, \citenamefont {Milder}, \citenamefont {Yin},
		\citenamefont {Palastro},\ and\ \citenamefont {Froula}}]{Hansen22}%
	\BibitemOpen
	\bibfield  {author} {\bibinfo {author} {\bibfnamefont {A.~M.}\ \bibnamefont
			{Hansen}}, \bibinfo {author} {\bibfnamefont {K.~L.}\ \bibnamefont {Nguyen}},
		\bibinfo {author} {\bibfnamefont {D.}~\bibnamefont {Turnbull}}, \bibinfo
		{author} {\bibfnamefont {B.~J.}\ \bibnamefont {Albright}}, \bibinfo {author}
		{\bibfnamefont {R.~K.}\ \bibnamefont {Follett}}, \bibinfo {author}
		{\bibfnamefont {R.}~\bibnamefont {Huff}}, \bibinfo {author} {\bibfnamefont
			{J.}~\bibnamefont {Katz}}, \bibinfo {author} {\bibfnamefont {D.}~\bibnamefont
			{Mastrosimone}}, \bibinfo {author} {\bibfnamefont {A.~L.}\ \bibnamefont
			{Milder}}, \bibinfo {author} {\bibfnamefont {L.}~\bibnamefont {Yin}},
		\bibinfo {author} {\bibfnamefont {J.~P.}\ \bibnamefont {Palastro}},\ and\
		\bibinfo {author} {\bibfnamefont {D.~H.}\ \bibnamefont {Froula}},\ }\bibfield
	{title} {\bibinfo {title} {Cross-beam energy transfer saturation: ion
			heating and pump depletion},\ }\href
	{https://doi.org/10.1088/1361-6587/ac493b} {\bibfield  {journal} {\bibinfo
			{journal} {Plasma Physics and Controlled Fusion}\ }\textbf {\bibinfo {volume}
			{64}},\ \bibinfo {pages} {034003} (\bibinfo {year} {2022})}\BibitemShut
	{NoStop}%
	\bibitem [{\citenamefont {Seaton}\ \emph {et~al.}(2022)\citenamefont {Seaton},
		\citenamefont {Yin}, \citenamefont {Follett}, \citenamefont {Albright},\ and\
		\citenamefont {Le}}]{Seaton22}%
	\BibitemOpen
	\bibfield  {author} {\bibinfo {author} {\bibfnamefont {A.~G.}\ \bibnamefont
			{Seaton}}, \bibinfo {author} {\bibfnamefont {L.}~\bibnamefont {Yin}},
		\bibinfo {author} {\bibfnamefont {R.~K.}\ \bibnamefont {Follett}}, \bibinfo
		{author} {\bibfnamefont {B.~J.}\ \bibnamefont {Albright}},\ and\ \bibinfo
		{author} {\bibfnamefont {A.}~\bibnamefont {Le}},\ }\bibfield  {title}
	{\bibinfo {title} {{Cross-beam energy transfer in direct-drive ICF. I.
				Nonlinear and kinetic effects}},\ }\href {https://doi.org/10.1063/5.0078800}
	{\bibfield  {journal} {\bibinfo  {journal} {Physics of Plasmas}\ }\textbf
		{\bibinfo {volume} {29}},\ \bibinfo {pages} {042706} (\bibinfo {year}
		{2022})}\BibitemShut {NoStop}%
	\bibitem [{\citenamefont {Yin}\ \emph {et~al.}(2023)\citenamefont {Yin},
		\citenamefont {Nguyen}, \citenamefont {Chen}, \citenamefont {Chacon},
		\citenamefont {Stark}, \citenamefont {Green},\ and\ \citenamefont
		{Haines}}]{Yin23}%
	\BibitemOpen
	\bibfield  {author} {\bibinfo {author} {\bibfnamefont {L.}~\bibnamefont
			{Yin}}, \bibinfo {author} {\bibfnamefont {T.~B.}\ \bibnamefont {Nguyen}},
		\bibinfo {author} {\bibfnamefont {G.}~\bibnamefont {Chen}}, \bibinfo {author}
		{\bibfnamefont {L.}~\bibnamefont {Chacon}}, \bibinfo {author} {\bibfnamefont
			{D.~J.}\ \bibnamefont {Stark}}, \bibinfo {author} {\bibfnamefont
			{L.}~\bibnamefont {Green}},\ and\ \bibinfo {author} {\bibfnamefont {B.~M.}\
			\bibnamefont {Haines}},\ }\bibfield  {title} {\bibinfo {title}
		{Time-dependent saturation and physics-based nonlinear model of cross-beam
			energy transfer},\ }\href {https://doi.org/10.1063/5.0134867} {\bibfield
		{journal} {\bibinfo  {journal} {Physics of Plasmas}\ }\textbf {\bibinfo
			{volume} {30}},\ \bibinfo {pages} {042706} (\bibinfo {year}
		{2023})}\BibitemShut {NoStop}%
	\bibitem [{\citenamefont {Shi}(2025)}]{Shi25data}%
	\BibitemOpen
	\bibfield  {author} {\bibinfo {author} {\bibfnamefont {Y.}~\bibnamefont
			{Shi}},\ }\href@noop {} {\bibinfo {title} {Particle-in-cell simulations of
			magnetized crossbeam energy transfer [data set]}},\ \bibinfo {howpublished}
	{Zenodo. \url{https://doi.org/10.5281/zenodo.16498564}} (\bibinfo {year}
	{2025})\BibitemShut {NoStop}%
	\bibitem [{\citenamefont {Taflove}\ \emph {et~al.}(2005)\citenamefont
		{Taflove}, \citenamefont {Hagness},\ and\ \citenamefont
		{Piket-May}}]{Taflove05}%
	\BibitemOpen
	\bibfield  {author} {\bibinfo {author} {\bibfnamefont {A.}~\bibnamefont
			{Taflove}}, \bibinfo {author} {\bibfnamefont {S.~C.}\ \bibnamefont
			{Hagness}},\ and\ \bibinfo {author} {\bibfnamefont {M.}~\bibnamefont
			{Piket-May}},\ }\bibfield  {title} {\bibinfo {title} {Computational
			electromagnetics: the finite-difference time-domain method},\ }\href@noop {}
	{\bibfield  {journal} {\bibinfo  {journal} {The Electrical Engineering
				Handbook}\ }\textbf {\bibinfo {volume} {3}},\ \bibinfo {pages} {15} (\bibinfo
		{year} {2005})}\BibitemShut {NoStop}%
	\bibitem [{\citenamefont {Shi}(2019)}]{PhysRevE.99.063212}%
	\BibitemOpen
	\bibfield  {author} {\bibinfo {author} {\bibfnamefont {Y.}~\bibnamefont
			{Shi}},\ }\bibfield  {title} {\bibinfo {title} {Three-wave interactions in
			magnetized warm-fluid plasmas: General theory with evaluable coupling
			coefficient},\ }\href {https://doi.org/10.1103/PhysRevE.99.063212} {\bibfield
		{journal} {\bibinfo  {journal} {Physical Review E}\ }\textbf {\bibinfo
			{volume} {99}},\ \bibinfo {pages} {063212} (\bibinfo {year}
		{2019})}\BibitemShut {NoStop}%
	\bibitem [{\citenamefont {Shi}(2022{\natexlab{a}})}]{Shi22}%
	\BibitemOpen
	\bibfield  {author} {\bibinfo {author} {\bibfnamefont {Y.}~\bibnamefont
			{Shi}},\ }\href@noop {} {\bibinfo {title} {{MATLAB} code for evaluating
			three-wave coupling in magnetized warm-fluid plasmas}},\ \bibinfo
	{howpublished} {\url{https://gitlab.com/seanYuanSHI/three-wave-matlab}}
	(\bibinfo {year} {2022}{\natexlab{a}})\BibitemShut {NoStop}%
	\bibitem [{\citenamefont {Shi}(2022{\natexlab{b}})}]{Shi25code}%
	\BibitemOpen
	\bibfield  {author} {\bibinfo {author} {\bibfnamefont {Y.}~\bibnamefont
			{Shi}},\ }\href@noop {} {\bibinfo {title} {{MATLAB} code for analyzing
			particle-in-cell simulations for magnetized crossbeam energy transfer}},\
	\bibinfo {howpublished}
	{\url{https://gitlab.com/seanYuanSHI/magnetized-cross-beam-energy-transfer}}
	(\bibinfo {year} {2022}{\natexlab{b}})\BibitemShut {NoStop}%
	\bibitem [{\citenamefont {Shi}\ \emph {et~al.}(2017)\citenamefont {Shi},
		\citenamefont {Qin},\ and\ \citenamefont {Fisch}}]{PhysRevE.96.023204}%
	\BibitemOpen
	\bibfield  {author} {\bibinfo {author} {\bibfnamefont {Y.}~\bibnamefont
			{Shi}}, \bibinfo {author} {\bibfnamefont {H.}~\bibnamefont {Qin}},\ and\
		\bibinfo {author} {\bibfnamefont {N.~J.}\ \bibnamefont {Fisch}},\ }\bibfield
	{title} {\bibinfo {title} {Three-wave scattering in magnetized plasmas: From
			cold fluid to quantized lagrangian},\ }\href
	{https://doi.org/10.1103/PhysRevE.96.023204} {\bibfield  {journal} {\bibinfo
			{journal} {Physical Review E}\ }\textbf {\bibinfo {volume} {96}},\ \bibinfo
		{pages} {023204} (\bibinfo {year} {2017})}\BibitemShut {NoStop}%
	\bibitem [{\citenamefont {Shi}\ and\ \citenamefont
		{Fisch}(2019)}]{10.1063/1.5099513}%
	\BibitemOpen
	\bibfield  {author} {\bibinfo {author} {\bibfnamefont {Y.}~\bibnamefont
			{Shi}}\ and\ \bibinfo {author} {\bibfnamefont {N.~J.}\ \bibnamefont
			{Fisch}},\ }\bibfield  {title} {\bibinfo {title} {Amplification of
			mid-infrared lasers via backscattering in magnetized plasmas},\ }\href
	{https://doi.org/10.1063/1.5099513} {\bibfield  {journal} {\bibinfo
			{journal} {Physics of Plasmas}\ }\textbf {\bibinfo {volume} {26}},\ \bibinfo
		{pages} {072114} (\bibinfo {year} {2019})}\BibitemShut {NoStop}%
\end{thebibliography}
%

\end{document}